\documentclass[%
twocolumn,
 amsmath,amssymb,
zpra,
]{revtex4-1}

\usepackage{graphicx,epsfig,epstopdf}
\usepackage{amsmath}
\usepackage{color}
\usepackage{caption}
\usepackage{subfigure}
\usepackage{array}
\usepackage{tabularx}
\usepackage{multirow}
\newcolumntype{L}[1]{>{\raggedright\arraybackslash}p{#1}}
\newcolumntype{C}[1]{>{\centering\arraybackslash}p{#1}}
\newcolumntype{M}[1]{>{\centering\arraybackslash}m{#1}}

\def\e{\begin{equation}}
\def\f{\end{equation}}
\def\_#1{{\bf #1}}
\def\.{\cdot}

\def\Re{{\rm Re\mit}}
\def\Im{{\rm Im\mit}}

\begin{document}

\title{Acoustic metasurfaces for scattering-free anomalous reflection and refraction}

\author{
 A.~D\'{i}az-Rubio$^{1}$  and   S.~A.~Tretyakov$^{1}$
}
 
\affiliation{$^1$Department of Electronics and Nanoengineering, Aalto University, P.~O.~Box~15500, FI-00076 Aalto, Finland}

\begin{abstract}

Manipulation of acoustic wavefronts by thin and planar devices, known as metasurfaces, has been extensively studied, in view of many important applications. 
Reflective and refractive metasurfaces are designed using the generalized reflection and Snell's laws, which tell that local phase shifts at the metasurface supply extra momentum to the wave, presumably allowing arbitrary control of reflected or transmitted waves. 
However, as it has been recently shown for the electromagnetic counterpart, conventional metasurfaces based on the generalized laws of reflection and refraction have important drawbacks in terms of power efficiency.
This work presents a new synthesis method of acoustic metasurfaces for anomalous reflection and transmission that overcomes the fundamental limitations of conventional designs, allowing full control of acoustic energy flow.
The results show that different mechanisms are necessary in the reflection and transmission scenarios for ensuring perfect performance.
Metasurfaces for anomalous reflection require non-local response, which allows energy channeling along the metasurface. On other hand, for perfect manipulation of anomalously transmitted waves, local and non-symmetric response is required.  
These conclusions are interpreted through appropriate surface impedance models which are used to find possible physical implementations of perfect metasurfaces in each scenario. 
We hope that this advance in the design of acoustic metasurfaces opens new avenues not only for perfect anomalous reflection and transmission but also for realizing more complex functionalities, such as focusing, self-bending or vortex generation.  
  
\end{abstract}

\maketitle

\section{Introduction}

The interest in quasi two-dimensional devices capable of manipulating waves revived with the formulation of the generalized reflection and Snell's laws \cite{GSL}, which shows a possibility of tailoring the direction of reflected and transmitted waves by introducing gradient phase shifts at the interface between two media. 
The generalized laws of reflection and refraction have been applied for controlling the direction of transmitted and reflected waves in electromagnetism \cite{Snell1}  and acoustics \cite{review,Snell3,Snell4,Snell5,Snell6,Snell7,Tx1,Tx2,Tx3, Tx4, Tx5}. 
By appropriately varying the phase shift introduced along the metasurface between $0$ and $2\pi$ the propagation direction of the reflected/refracted wave can be controlled. These approaches enable tailoring the energy propagation direction, but with important restrictions of the efficiency (the amount of energy that is sent into the desired direction is smaller than the energy introduced in the system, even for lossless metasurfaces).

Recently, it has been demonstrated that some additional considerations about the power conservation can be applied over the conventional generalized reflection and Snell's laws for ensuring full control of electromagnetic energy flow \cite{synthesis,synthesis_elef,Alu,last}. 
This advance has attracted much attention due to the possibility of dramatic improvements of conventional solutions, especially for steep reflection or transmission angles. 
For electromagnetic waves, the  basis of this ``second generation'' of gradient electromagnetic metasurfaces has been established and numerically verified, and for reflective metasurfaces  the theoretical findings have been already confirmed experimentally \cite{Reflectarray_patches}. 

However, it appears that the synthesis tools for acoustic metasurfaces do not benefit from the new knowledge. 
In the acoustic reflection scenario, the generalized reflection law has been experimental demonstrated \cite{Snell3,Snell4,Snell5,Snell6,Snell7} using labyrinthine unit cells which provide a phase-shift profile with the $2\pi$ span in the reflection coefficient phase. 
However, in view of the results of \cite{synthesis,synthesis_elef,Alu,last}, the performance of these designs is not optimal because significant energy is spread in unwanted directions. Theoretical studies based of inhomogeneous impedance along the metasurface \cite{Snell4} show the coexistence of more than one reflected wave. 
For perfect control of anomalous reflection or, in other words, for allowing arbitrary changes of the direction of reflected plane waves, we need to ensure perfect steering of all the incident power into the desired direction avoiding generation of parasitic waves propagating in other directions or losses in the system.  

On the other hand, the same approach based on the generalized Snell's law has been applied for the design of refractive acoustic metasurfaces \cite{Tx1,Tx2,Tx3,Tx4,Tx5}. 
The direction of the transmitted wave is controlled by linearly modulating the local phase shift in transmission through the metasurface. 
For the design of unit cells, different topologies  have been used, including space-colling structures \cite{Tx1}, slits filled with different density materials \cite{Tx3} or straight pipes with Helmholtz resonators in series \cite{Tx2,Tx4}. 
One problem addressed in the design of refractive metasurface was the required impedance matching of each meta-atom in order to obtain total transmission. 
In this sense,  substantial improvements in the design of matched unit cell have been achieved by using tapered labyrinthine units \cite{Tx5}.
However, as it has been  recently demonstrated for the electromagnetic scenario \cite{synthesis,synthesis_elef,Review_elef}, by ensuring perfect matching in the microscopic design of the metasurface (individual design of each meta-atom) we cannot obtain the proper macroscopic behavior of the metasurface.

In this work we present the foundations for the synthesis of perfect acoustic metasurfaces, overcoming the fundamental limitations of conventional designs. 
The study covers two different scenarios: anomalous reflection and transmission of acoustic plane waves. With the purpose of simplifying the presentation and emphasizing the novelty of this approach, in both cases the analysis starts with a comprehensive overview of  the known  approaches based on the generalized reflection and Snell's laws. 
After identifying the weaknesses of current designs, we propose new methods which ensure perfect control of acoustic energy in reflection and refraction. 
Finally, we interpret the theoretical findings in terms of the physical properties of metasurface unit cells and give examples of possible realizations.

\section{Acoustic metasurfaces for reflection}

In this section, the reflected wavefront manipulation is studied. Particularly, we focus the study on  anomalous reflection of acoustic plane waves.   This fundamental functionality is the base of many interesting applications such reflection lenses, plane wave to surface wave conversion or acoustic retro-reflectors.  

\subsection{Design based on the generalized reflection law}

\begin{figure}[h!]
	\centering
	\subfigure[]{\includegraphics[width=0.55\linewidth]{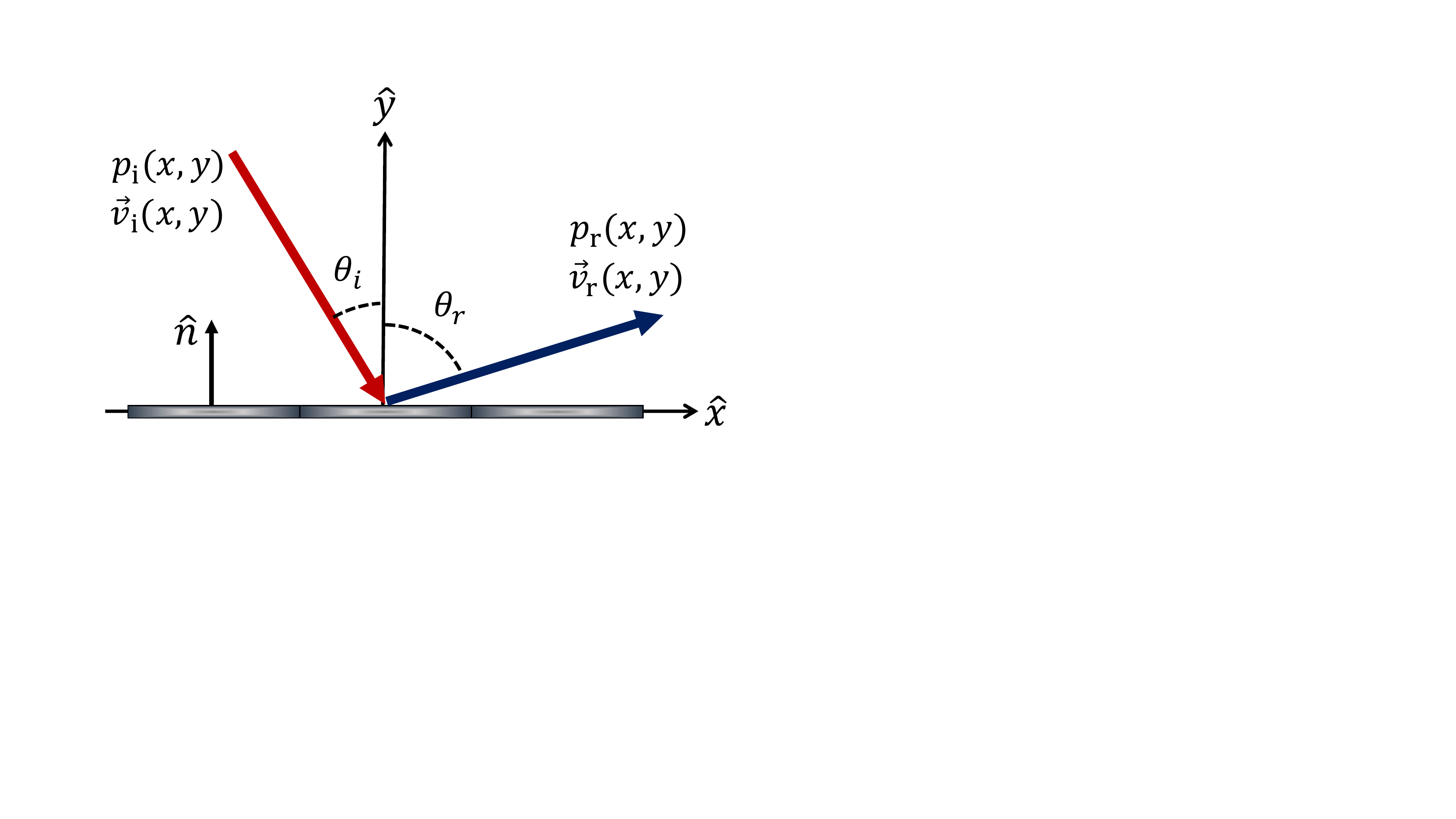}\label{fig:Schematic_Rx_A}}
	\subfigure[]{\includegraphics[width=0.3\linewidth]{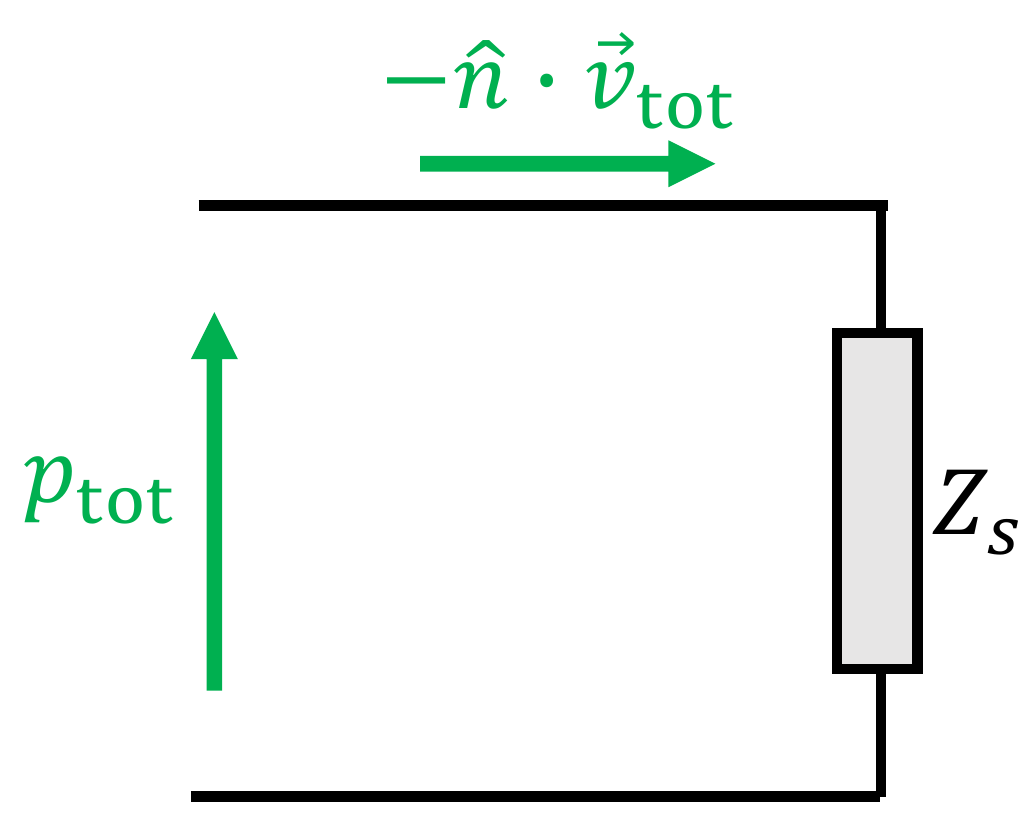}\label{fig:Schematic_Rx_B}}
	\caption{(a) Schematic representations of the desired metasurface behavior for the anomalous  reflection scenario.  (b) Equivalent circuit proposed for the analysis of reflective metasurfaces. }\label{fig:Schematic_Rx}
\end{figure}

In order to understand the current status of the synthesis methods, we start with the analysis of reflective gradient metasurfaces based on the generalized reflection law (we use the same short-hand notation, GSL, for both generalized Snell's law and the generalized reflection law). If we consider the scenario illustrated in Fig.~\ref{fig:Schematic_Rx_A} where the incident and reflected waves propagate in a homogeneous medium with density $\rho$ and sound speed $c$, assuming time-harmonic dependence  $e^{j\omega t}$, the incident and reflected pressure fields can be written as
\begin{eqnarray}
p_{\rm{i}}(x,y)=p_0e^{-jk\sin\theta_{\rm i} x}e^{jk\cos\theta_{\rm i} y}, \\
p_{\rm{r}}(x,y)=Ap_{\rm 0}e^{-jk\sin\theta_{\rm r} x}e^{-jk\cos\theta_{\rm r} y},
\end{eqnarray}
where $p_0$ is the amplitude of the incident plane wave, $k=\omega/c$ is the wavenumber in the background medium at the operation frequency, $\theta_{\rm i}$ and $\theta_{\rm r}$ are the incidence and reflection angles, and $A$ is a constant which relates the amplitudes of the incident and reflected waves. The velocity vectors associated with these pressure fields ($\vec{v}=\frac{j}{\omega\rho}\nabla p$) read
 \begin{eqnarray}
\vec{v}_{\rm i}(x,y)=\frac{p_{\rm{i}}(x,y)}{Z_0}\left(\sin{\theta_{\rm i}}\hat{x}-\cos{\theta_{\rm i}}\hat{y}\right),\\
 \vec{v}_{\rm r}(x,y)=\frac{p_{\rm{r}}(x,y)}{Z_0}\left(\sin{\theta_{\rm r}}\hat{x} +\cos{\theta_{\rm r}}\hat{y}\right),\end{eqnarray} 
where $Z_{\rm 0}=c\rho$ is the characteristic impedance of the background medium.

Assuming  that the field beyond the metasurface is zero (an impenetrable metasurface), the system can be conveniently modeled by the equivalent circuit shown in  Fig.~\ref{fig:Schematic_Rx_B} where the impedance $Z_{\rm s}$ models the specific impedance of the metasurface. GSL designs are based on the assumption that a linear gradient phase shift [$\frac{\partial \Phi_x}{\partial x}=k(\sin\theta_{\rm r}-\sin\theta_{\rm i}) $] is introduced by the metasurface. In other words, the metasurface is characterized by the  local reflection coefficient with the unit amplitude, which can be written as 
\e \Gamma(x)=\frac{e^{-jk\sin\theta_{\rm r}x}}{e^{-jk\sin\theta_{\rm i}x}}=e^{jk(\sin\theta_{\rm i}-\sin\theta_{\rm r})x}=e^{j\Phi_x},\f
where the reflection phase is defined as $\Phi_x=k(\sin\theta_{\rm r}-\sin\theta_{\rm i})x$. The reflection coefficient is related with the surface impedance as $\Gamma=\frac{Z_{\rm s}-Z_{\rm i}}{Z_{\rm s}+Z_{\rm i}}$, where $Z_{\rm i}=Z_0/\cos\theta_{\rm i}$ represents the specific acoustic impedance of the incident wave at the metasurface. From this expression, the  impedance which models the metasurface can be found as
\e Z_{\rm s}(x)=j\frac{Z_0}{\cos\theta_{\rm i}}\cot(\Phi_x/2)\label{eq:ImpRX_conv}.\f

\begin{figure}[h!]
	\centering
	\begin{minipage}{1\columnwidth}
	\subfigure[]{\includegraphics[width=0.57\linewidth]{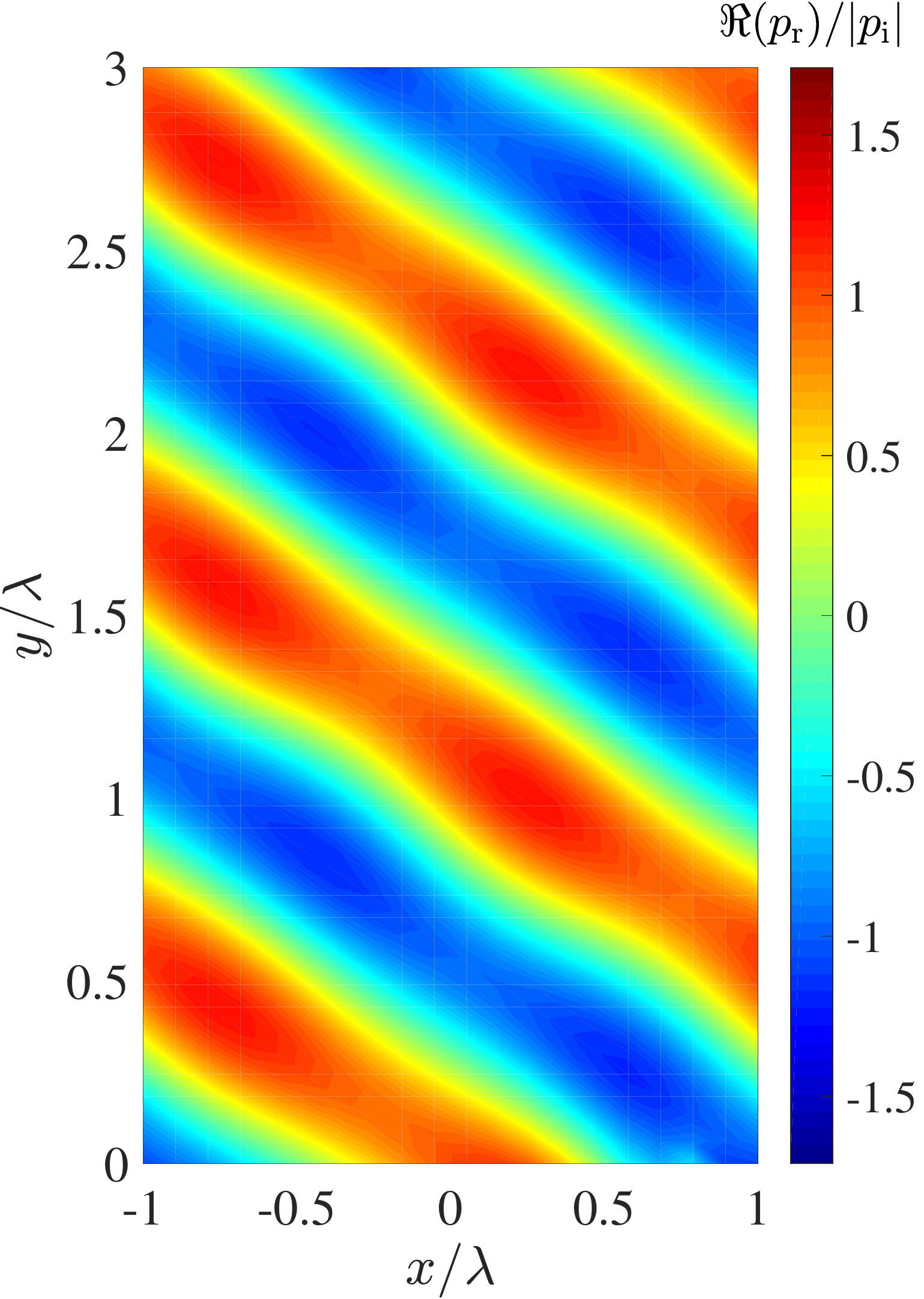}\label{fig:RX_conventional_A}}
	\subfigure[]{\includegraphics[width=0.36\linewidth]{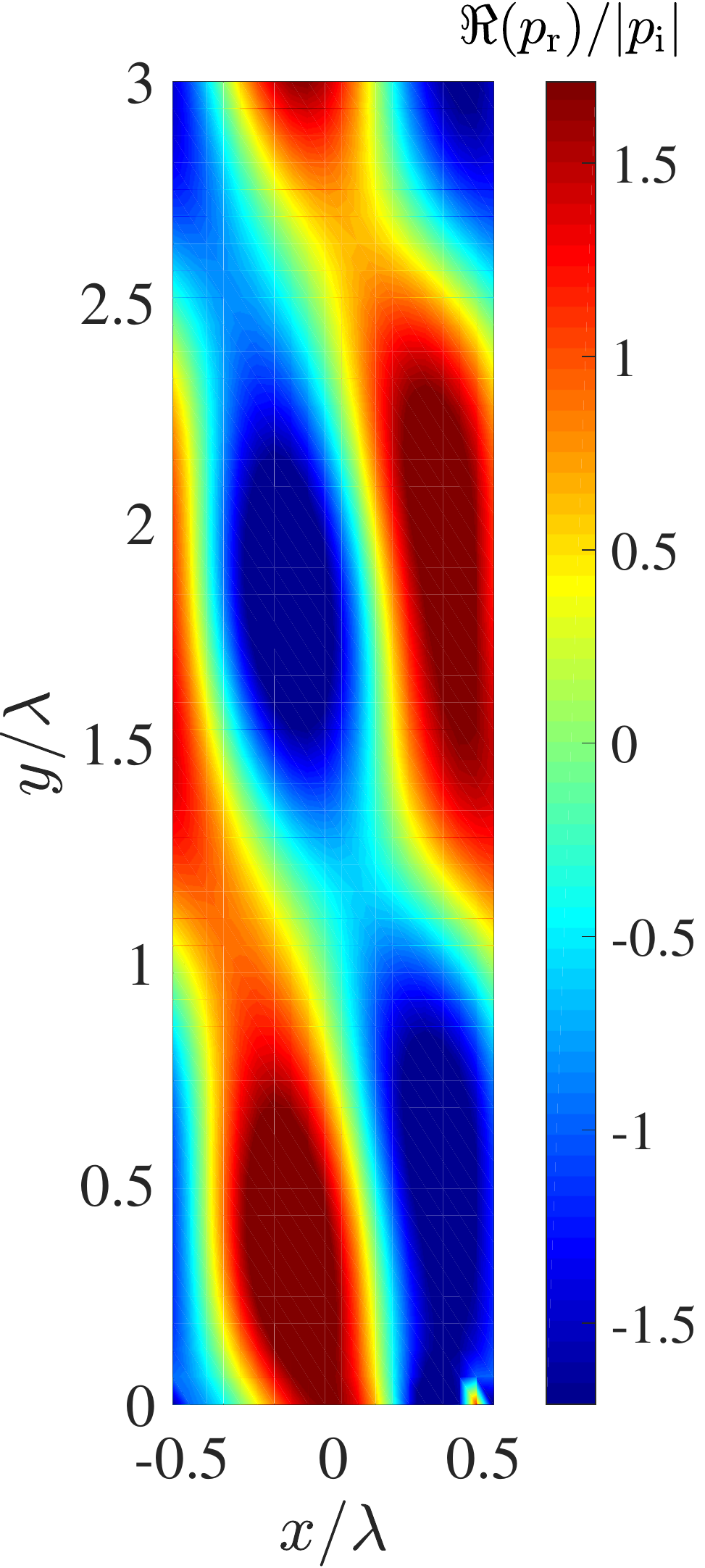}\label{fig:RX_conventional_B}}
	\end{minipage}
	\begin{minipage}{1\columnwidth}
	\subfigure[]{\includegraphics[width=0.47\linewidth]{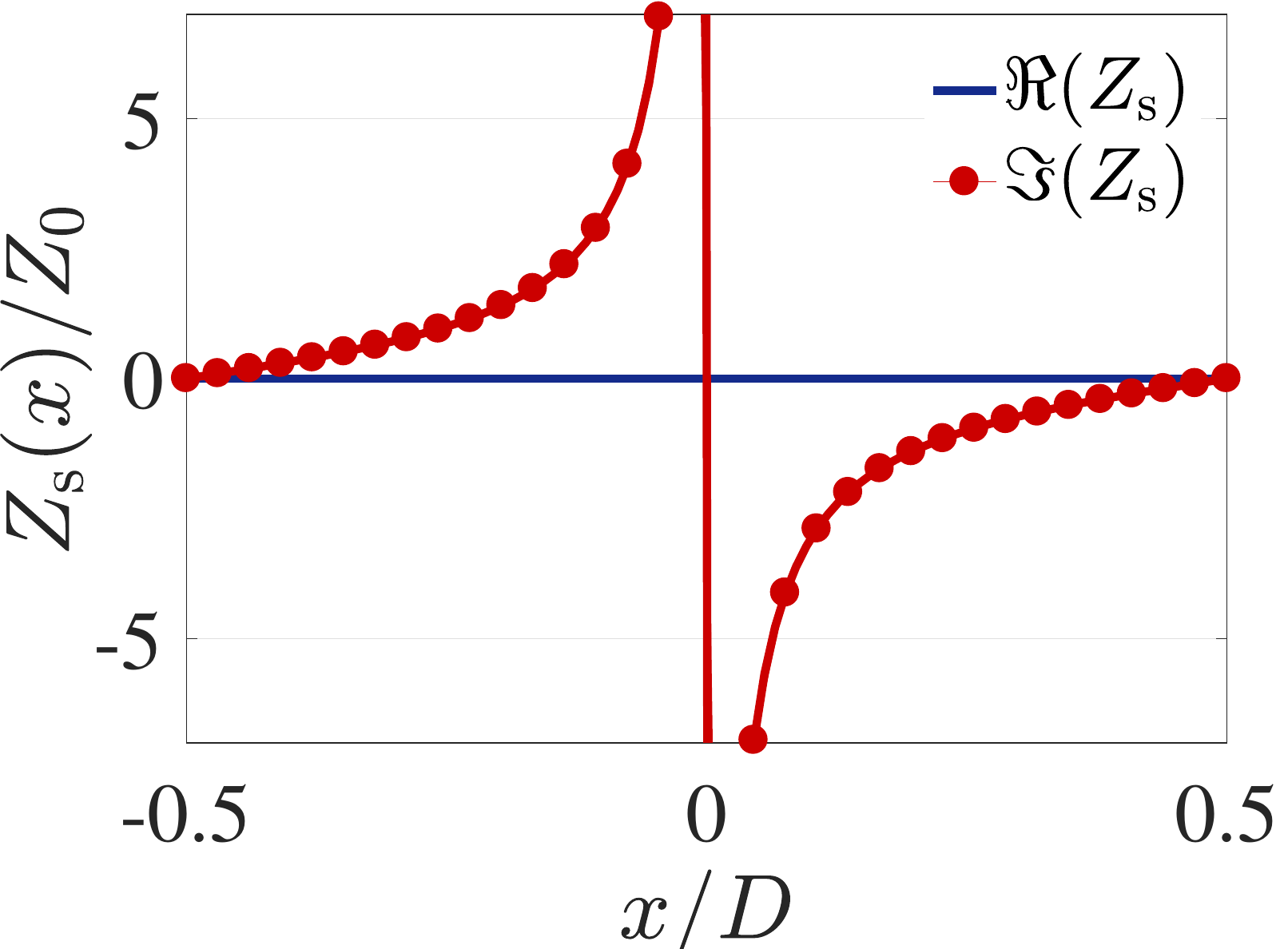}\label{fig:RX_conventional_C}}
	\subfigure[]{\includegraphics[width=0.45\linewidth]{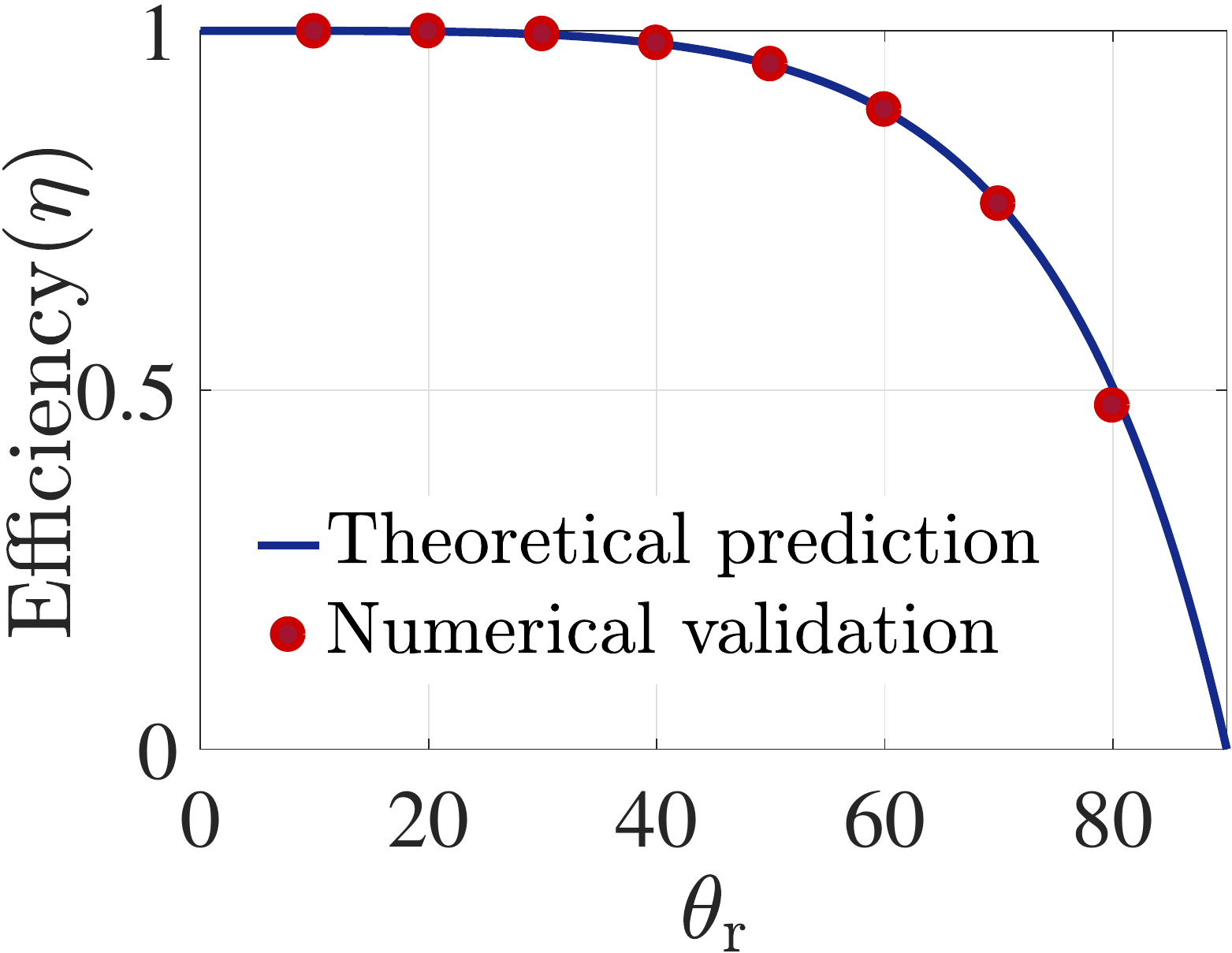}\label{fig:RX_conventional_D}}
	\end{minipage}
	\caption{Real part of the scattered pressure field for a metasurface designed according to  Eq.~(\ref{eq:ImpRX_conv}) when:  (a) $\theta_{\rm i}=0^\circ$ and $\theta_{\rm r}=30^\circ$;  (b) $\theta_{\rm i}=0^\circ$ and $\theta_{\rm r}=70^\circ$.  (c) Surface impedance described by Eq.~(\ref{eq:ImpRX_conv}) when $\theta_{\rm i}=0^\circ$ and $\theta_{\rm r}=70^\circ$.  (d) Efficiency of the GSL gradient metasurfaces as a function of the reflection angle. }\label{fig:RX_conventional}
\end{figure}

Figure \ref{fig:RX_conventional} presents a numerical study of the conventional designs based on the GSL. The surface impedance modeled by  Eq.~(\ref{eq:ImpRX_conv}) is purely imaginary [see Fig.~\ref{fig:RX_conventional_C}], so lossless implementations are possible for this kind of reflective metasurfaces. Actual implementations of the desired impedance profiles can be obtained by using simple rigidly terminated waveguides with different lengths or, exploiting the longitudinal character of acoustics waves and so the absence of cutoff frequency, ``space-colling" particles with labyrinth channels \cite{Snell3,Snell4,Snell5,Snell6,Snell7}. In these cases, each meta-atom has to be carefully tailored for individually implementing the required surface impedance profile and producing the required local phase shift. For the purposes of this study  we assume that the required impedance profile has been realized and model the metasurface (using COMSOL software) as an impedance boundary described by  Eq.~(\ref{eq:ImpRX_conv}). Figures~\ref{fig:RX_conventional_A} and  \ref{fig:RX_conventional_B} show the results of numerical simulations when the metasurface is illuminated normally ($\theta_{\rm i}=0^\circ$) and the reflection angles are $\theta_{\rm r}=30^\circ$  and $\theta_{\rm r}=70^\circ$, respectively. From the comparison of these two examples, it is easy to see two important issues: First, the ``quality'' of the reflected wave decreases when the reflection angle increases, due to  parasitic reflections in other directions; Second, the amplitude of the reflected wave changes with the reflection angle, although this behavior it is not contemplated in the design statement ($A=1$). Clearly, the simple design philosophy described by Eq.~(\ref{eq:ImpRX_conv}) does not ensure the perfect conversion of energy between the incident and reflected plane waves and it cannot be considered as an accurate method for the design of metasurfaces for large values of the reflection angle. 

The conclusions extracted from the analysis of the numerical simulations can be understood as an impedance mismatch problem. Although the metasurface provides the desired phase response, the  incident and reflected waves have different specific impedances, so part of the energy  cannot be redirected into the desired direction. Since the metasurface is assumed to be lossless, part of the incident energy has to be reflected into other directions (into $0^\circ$ and $-70^\circ$ in the example of Fig.~\ref{fig:RX_conventional}). The reflections into parasitic directions can be estimated introducing reflection coefficient calculated in terms of the respective impedances:
\e R=\frac{Z_{\rm r}-Z_{\rm i}}{Z_{\rm r}+Z_{\rm i}}=\frac{\cos\theta_{\rm i}-\cos\theta_{\rm r}}{\cos\theta_{\rm i}+\cos\theta_{\rm r}}.\f
Because the metasurface is an impenetrable boundary, the total pressure of the incident and reflected waves ($1+R$) is equal to the pressure of the wave redirected into the desired direction ($A_{\rm GSL}$), in analogy to transmission of electromagnetic waves through electric-current sheets  \cite{Sergei_book}:
\e A_{\rm GSL}=1+R=\frac{2\cos\theta_{\rm i}}{\cos\theta_{\rm i}+\cos\theta_{\rm r}}.\label{A_GSL}\f

We can now define the efficiency of the metasurface  as the ratio between the incident power and the power reflected in the desired direction. The acoustic power can be expressed in terms of  the intensity vector
\begin{equation}
\vec{I}=\frac{1}{2}\Re{[p\vec{v}^*]},
\end{equation}
where '$*$' represents the complex conjugate. Due to the periodicity of the system only the normal component of the intensity vector will take part in the power balance. 
The normal component of the incident power is 
\begin{equation}
P_{\rm i}=\hat{n} \cdot\vec{I}_{\rm i}=-\frac{p_0^2}{Z_0}\cos{\theta_{\rm i}}.
\end{equation}
The normal component of the power carried in the  desired reflection direction can be calculated as
\begin{equation}
P_{\rm r}=\hat{n} \cdot\vec{I}_{\rm r}=A^2\frac{p_0^2}{Z_0}\cos{\theta_{\rm r}},
\end{equation}
and the efficiency reads
\begin{equation} \eta=\frac{|P_{\rm r}|}{|P_{\rm i}|}=A^2\frac{\cos\theta_{\rm r}}{\cos\theta_{\rm i}}.\label{eq:eff}\end{equation}
Substituting the wave amplitude $A$ from Eq. (\ref{A_GSL}) we can finally estimate the efficiency of conventional metasurfaces as
\begin{equation} \eta_{\rm GSL}=\left(\frac{2\cos\theta_{\rm i}}{\cos\theta_{\rm i}+\cos\theta_{\rm r}}\right)^2\frac{\cos\theta_{\rm r}}{\cos\theta_{\rm i}}.\label{eq:eff_GSL}\end{equation}

\begin{figure}[h!]
	\centering
	\begin{minipage}{1\columnwidth}
		\subfigure[]{\includegraphics[width=0.57\linewidth]{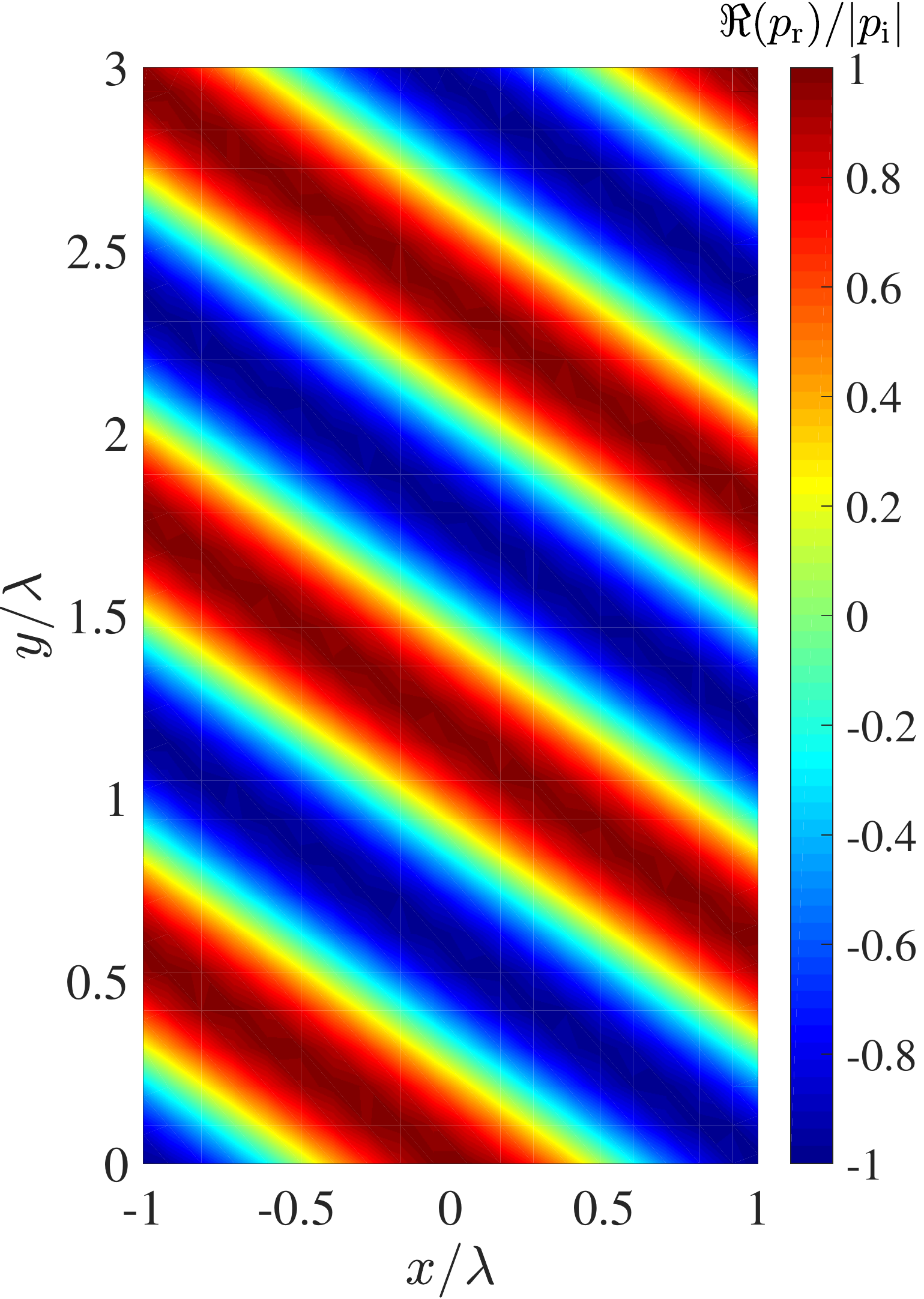}\label{fig:RX_lossy_A}}
		\subfigure[]{\includegraphics[width=0.36\linewidth]{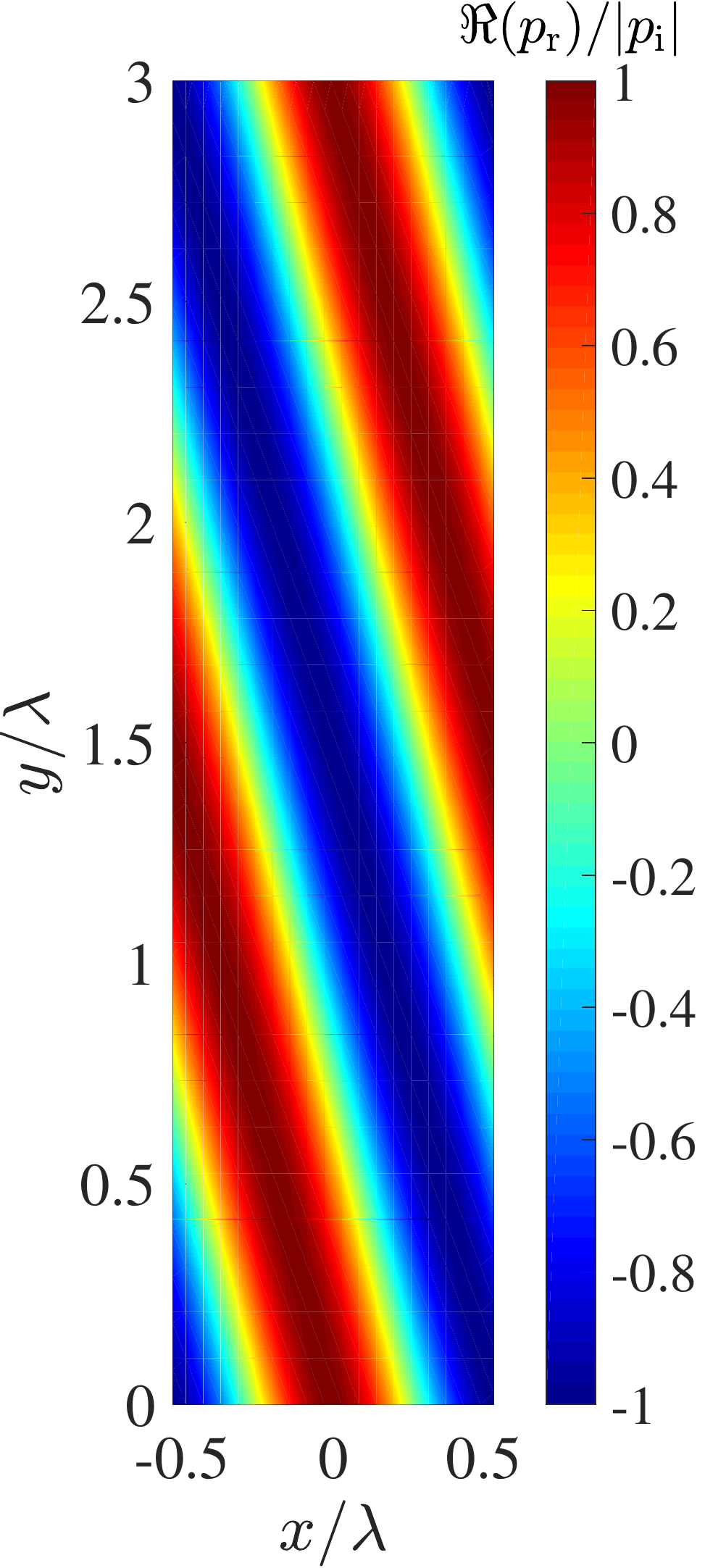}\label{fig:RX_lossy_B}}
	\end{minipage}
	\begin{minipage}{1\columnwidth}
		\subfigure[]{\includegraphics[width=0.47\linewidth]{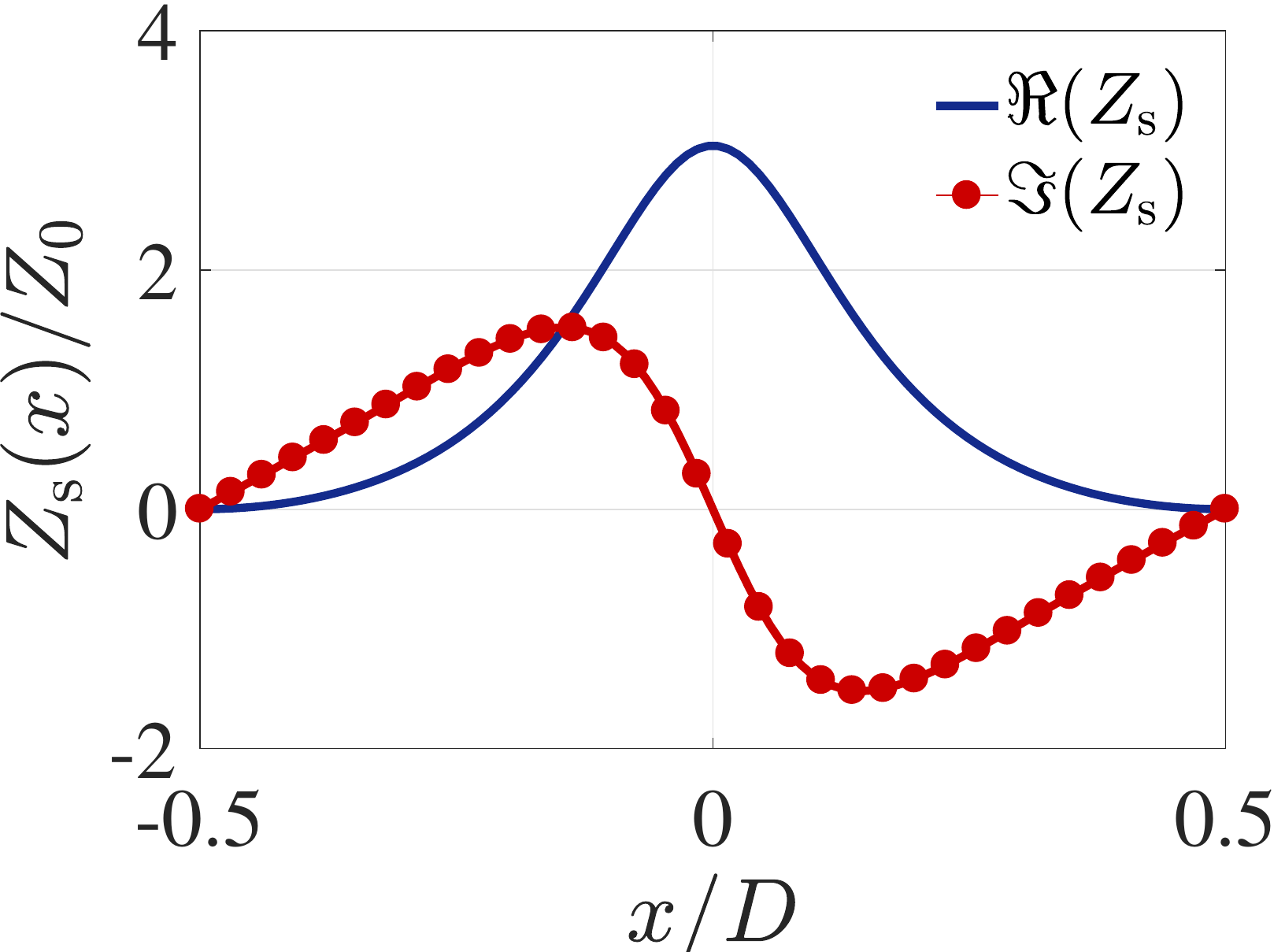}\label{fig:RX_lossy_C}}
		\subfigure[]{\includegraphics[width=0.45\linewidth]{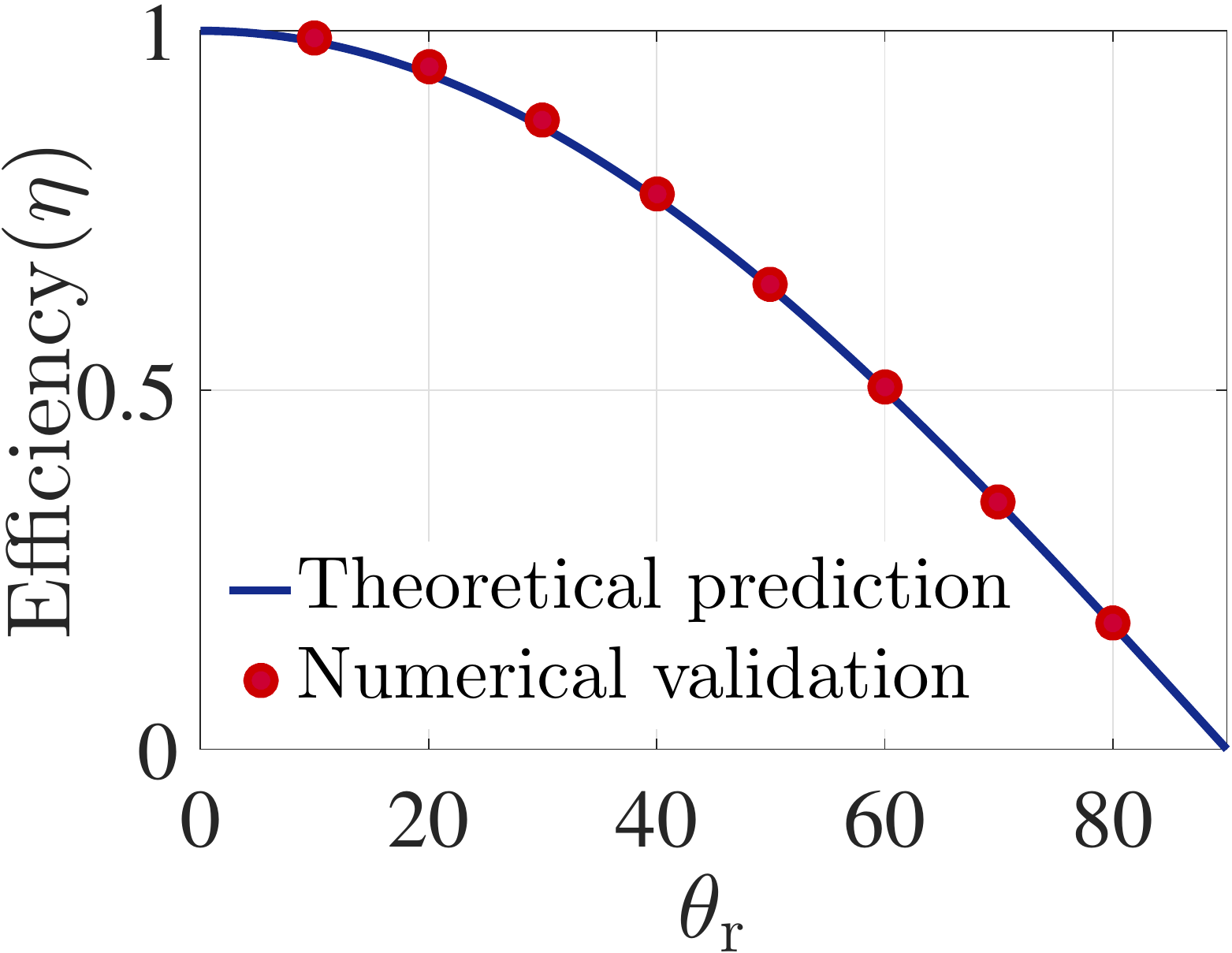}\label{fig:RX_lossy_D}}
	\end{minipage}
	\caption{Real part of the scattered pressure field for a metasurface designed according to  Eq.~(\ref{eq:rx_imp_lossy}) when:  (a) $\theta_{\rm i}=0^\circ$ and $\theta_{\rm r}=30^\circ$ and  (b) $\theta_{\rm i}=0^\circ$ and $\theta_{\rm r}=70^\circ$.  (c) Surface impedance described by Eq.~(\ref{eq:rx_imp_lossy}) when $\theta_{\rm i}=0^\circ$ and $\theta_{\rm r}=70^\circ$.  (d) Efficiency of the gradient metasurfaces for $A=1$ as a function of the reflection angle. }\label{fig:RX_lossy}
\end{figure}

Figure~\ref{fig:RX_conventional_D} represents the efficiency estimation given by Eq.~(\ref{eq:eff_GSL}) and its comparison with the numerical results. 
In order to find the power efficiency from numerical results we calculate the amplitude of the reflected plane wave into $\theta_{\rm r}$, $A_{\rm COMSOL}$. This amplitude can be calculated  as
\begin{equation} 
A_{\rm COMSOL}=\frac{1}{D}\int_{0}^{D} p_{\rm r} \. e^{jk\sin\theta_{\rm r} x}\, \label{A_comsol}dx,
\end{equation}
where $D$ is the metasurface period, and the  efficiency is obtained using Eq.~(\ref{eq:eff}). 
It is possible to see how the efficiency of the generalized reflection law metasurfaces dramatically decreases when the reflection angle increases.

\subsection{Lossy metasurfaces for anomalous reflection}

As we have seen, the amplitude of the reflected wave  in the conventional design is not equal to the amplitude of the incident plane wave ($A_{\rm GSL}\neq 1$).  If we design a metasurface which arbitrarily changes the direction of the reflected wave keeping the amplitude $A=1$, the pressure field at metasurface can be written as
\begin{equation} p_{\rm tot}(x,0)=p_0(1+e^{j\Phi_x})e^{-jk\sin\theta_{\rm i} x}.\label{pre_l}\end{equation}
The corresponding total velocity at the metasurface reads
\begin{eqnarray} \vec{v}_{\rm tot}(x,0)=\frac{p_0}{Z_0}(\sin\theta_{\rm i}+\sin\theta_{\rm r}e^{j\Phi_x})e^{-jk\sin\theta_{\rm i} x}\hat{x}+\\
\frac{p_0}{Z_0}(-\cos\theta_{\rm i}+\cos\theta_{\rm r}e^{j\Phi_x})e^{-jk\sin\theta_{\rm i} x}\hat{y}.\label{vel_l} \end{eqnarray}
At this point, we have to satisfy the boundary condition at the metasurface. We can do that by defining the surface impedance which models this metasurface as
\begin{equation}Z_{\rm s}(x)=\frac{p_{\rm tot}(x,0)}{-\hat{n} \cdot\vec{v}_{\rm tot}(x,0)}.\end{equation}
Introducing the expressions for the desired total pressure (\ref{pre_l}) and velocity (\ref{vel_l}) into this equation, we find the impedance which models such metasurfaces:
\begin{equation}Z_{\rm s}(x)=Z_0\frac{1+e^{j\Phi_x}}{\cos\theta_{\rm i}-\cos\theta_{\rm r}e^{j\Phi_x}}. \label{eq:rx_imp_lossy}\end{equation}

Figure~\ref{fig:RX_lossy} presents the results of a numerical study of acoustic metasurfaces based on  Eq.~(\ref{eq:rx_imp_lossy}). As in the previous case, Figs.~\ref{fig:RX_lossy_A} and  \ref{fig:RX_lossy_B} show simulated results for the  metasurface illuminated normally and designed for the reflection angles $\theta_{\rm r}=30^\circ$  and $\theta_{\rm r}=70^\circ$, respectively. The results confirm the required performance of metasurfaces which anomalously reflect a perfect plane have with the same amplitude as the incident wave. The surface impedance given by  Eq.~(\ref{eq:rx_imp_lossy}) is a complex number as it is shown in Fig.~\ref{fig:RX_lossy_C} for $\theta_{\rm i}=0^\circ$ and $\theta_{\rm r}=70^\circ$. The real part of the impedance takes positive values (modeling losses) over all the period showing that these metasurfaces are necessarily lossy, which is a condition for keeping the amplitude of the reflected wave equal to the incident wave. To illustrate this behavior we can analyze the efficiency of the metasurface found from Eq.~(\ref{eq:eff}) when $A=1$:
\begin{equation} \eta_{A=1}=\frac{\cos\theta_{\rm r}}{\cos\theta_{\rm i}}.\label{eq:eff_lossy}\end{equation}
This expression is represented in Fig.~\ref{fig:RX_lossy_D} as a function of the reflection angle for  $\theta_{\rm i}=0^\circ$. Numerical results have been calculated in the same way as before, using Eq.~(\ref{A_comsol}). We can see that the efficiency dramatically decreases when the reflection angle increases. When the reflection angle increases, the power
sent into the desired direction decreases and all the remaining energy is absorbed in the metasurface. 
On the other hand, if $\theta_{\rm i}>\theta_{\rm r}$ the real part of the surface impedance becomes negative (gain), meaning that additional energy has to be introduced in the system in other to obtain the desired performance.

\begin{figure}[]
	\centering
	\begin{minipage}{1\columnwidth}
		\subfigure[]{\includegraphics[width=0.57\linewidth]{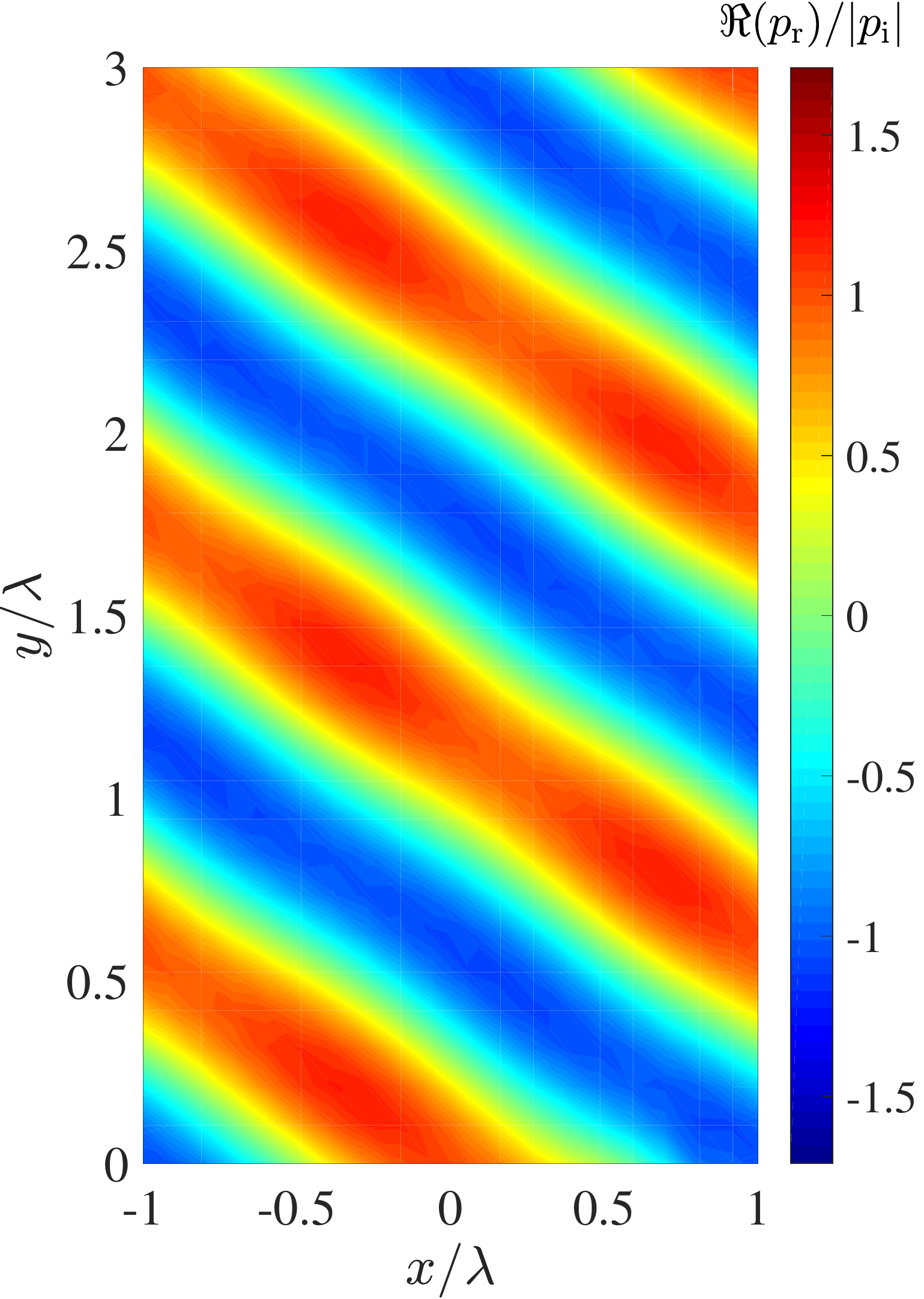}\label{fig:RX_active_A}}
		\subfigure[]{\includegraphics[width=0.36\linewidth]{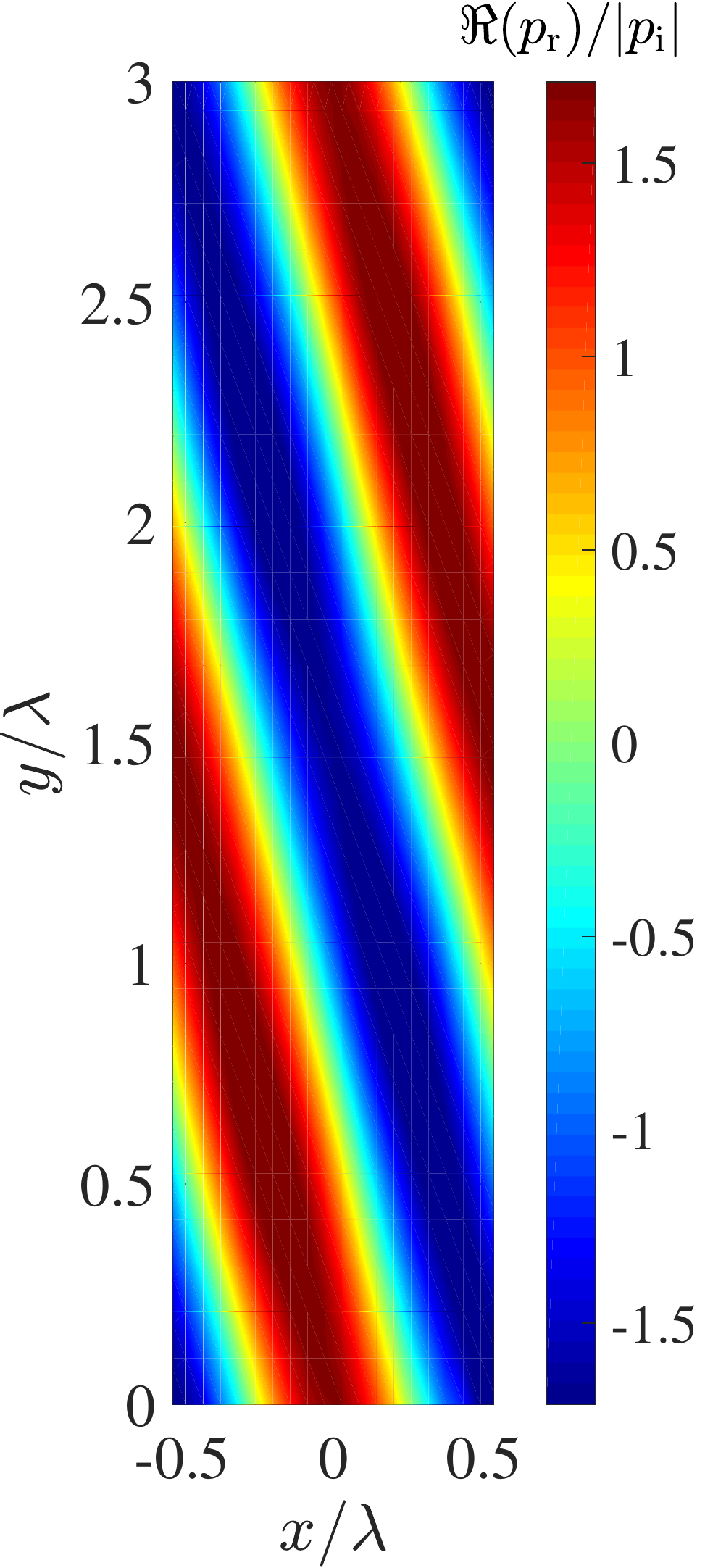}\label{fig:RX_active_B}}
	\end{minipage}
	\begin{minipage}{1\columnwidth}
		\subfigure[]{\includegraphics[width=0.47\linewidth]{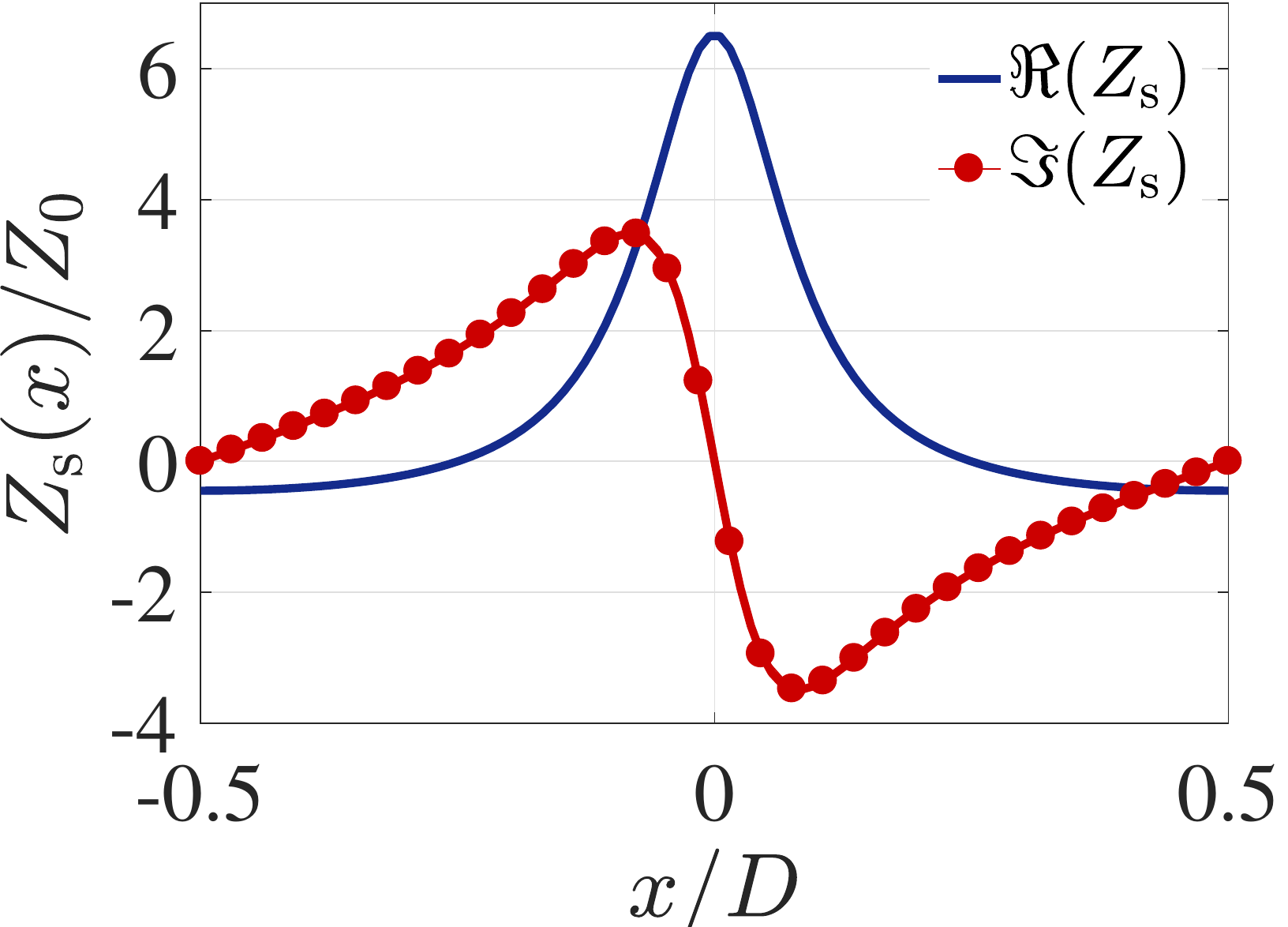}\label{fig:RX_active_C}}
		\subfigure[]{\includegraphics[width=0.45\linewidth]{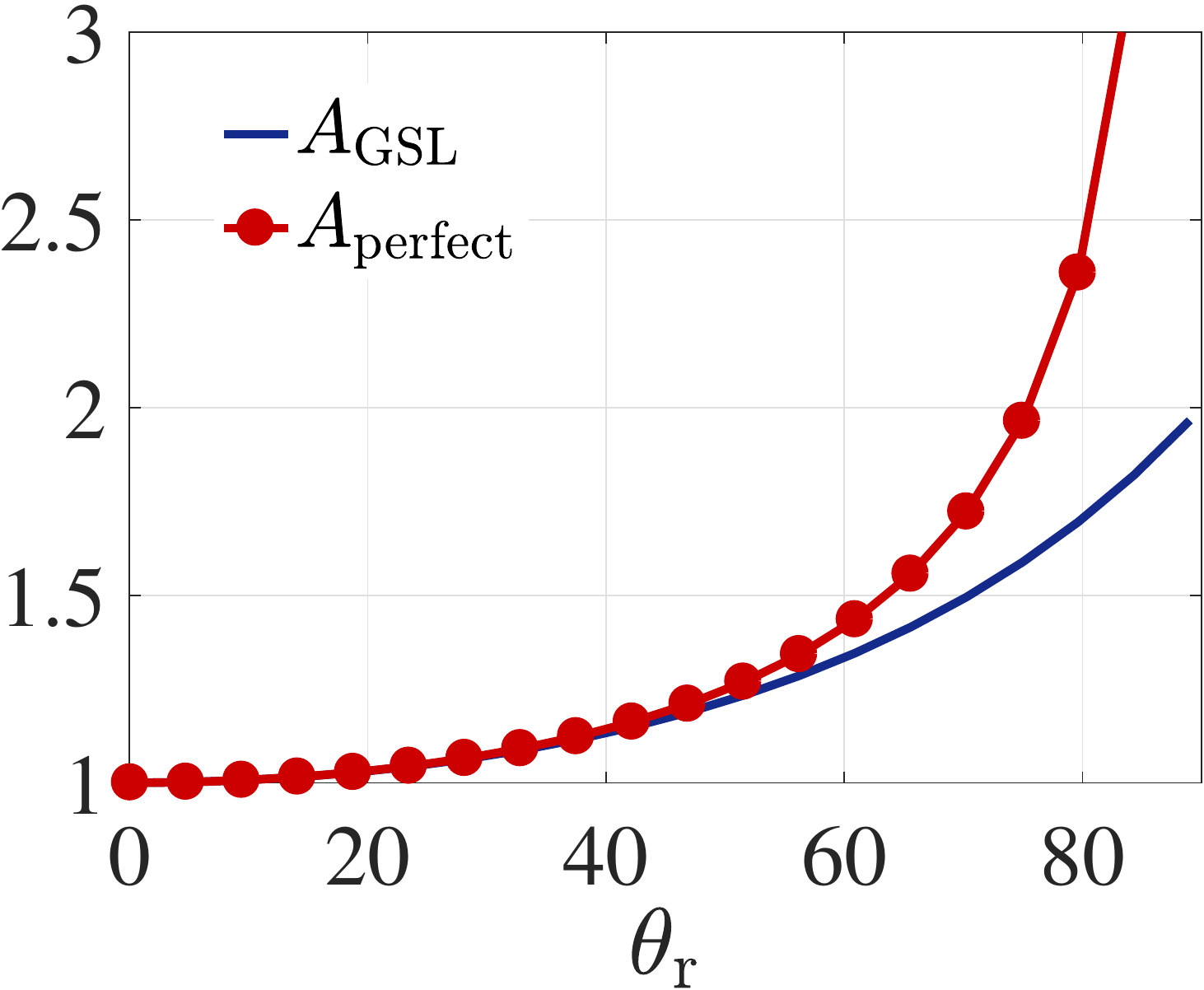}\label{fig:RX_active_D}}
	\end{minipage}
	\caption{Real part of the scattered pressure field for a metasurface designed according to  Eq.~(\ref{eq:rx_imp_active}) when:  (a) $\theta_{\rm i}=0^\circ$ and $\theta_{\rm r}=30^\circ$ and  (b) $\theta_{\rm i}=0^\circ$ and $\theta_{\rm r}=70^\circ$.  (c) Surface impedance described by Eq.~(\ref{eq:rx_imp_active}) when $\theta_{\rm i}=0^\circ$ and $\theta_{\rm r}=70^\circ$. (d) Amplitude of the reflected wave for perfect anomalous reflection (red symbols) compared with conventional designs based on GSL (blue line). }\label{fig:RX_active}
\end{figure}

\subsection{Active-lossy scenario and lossless non-local realization}

Obviously, the design approach defined by Eq.~(\ref{eq:rx_imp_lossy}) presents important drawbacks in terms of power efficiency, although there are no parasitic reflections into unwanted directions. For perfect anomalous reflection where all the impinging energy is sent into the desired direction we have to ensure $\eta=1$. From Eq.~(\ref{eq:eff}) it is easy to find the amplitude coefficient which corresponds to the perfect performance:  For perfect anomalous reflection the amplitude of the reflected wave has to be  $A=\sqrt{\cos\theta_{\rm i}/\cos\theta_{\rm r}}$. Figure \ref{fig:RX_active_D} shows a comparison between the required amplitude for perfect performance and the amplitude of conventional designs based on GSL when $\theta_{\rm i}=0^\circ$. The difference between both approaches increases with the angle of reflection, confirming our previous conclusion about the  poor efficiency of conventional design for large differences between incident and reflected angles. In this scenario, the acoustic impedance of the metasurface can be calculated writing the pressure field
\begin{equation}p_{\rm tot}(x,0)=p_0\left(1+\sqrt{\frac{\cos\theta_{\rm i}}{\cos\theta_{\rm r}}}e^{j\Phi_x}\right)e^{-jk\sin\theta_{\rm i} x}\end{equation}
and the velocity at the metasurface
\begin{equation}
\begin{split}\vec{v}_{\rm tot}(x,0)=\frac{p_0}{Z_0}\left(\sin\theta_{\rm i}+\sqrt{\frac{\cos\theta_{\rm i}}{\cos\theta_{\rm r}}}\sin\theta_{\rm r}e^{j\Phi_x}\right)e^{-jk\sin\theta_{\rm i} x}\hat{x}\\+\frac{p_0}{Z_0}\left(-\cos\theta_{\rm i}+\sqrt{\cos\theta_{\rm i}\cos\theta_{\rm r}}e^{j\Phi_x}\right)e^{-jk\sin\theta_{\rm i} x}\hat{y}.\end{split}\end{equation}
Finally, the corresponding surface impedance reads
\begin{equation} Z_{\rm s}(x)=\frac{Z_0}{\sqrt{\cos\theta_{\rm i}\cos\theta_{\rm r}}}
\frac{\sqrt{\cos\theta_{\rm r}}+\sqrt{\cos\theta_{\rm i}}e^{j\Phi_x}}{\sqrt{\cos\theta_{\rm i}}-\sqrt{\cos\theta_{\rm r}}e^{j\Phi_x}}.\label{eq:rx_imp_active}\end{equation}

In Fig.~\ref{fig:RX_active} the numerical results obtained with Eq.~(\ref{eq:rx_imp_active}) are represented.   Figures \ref{fig:RX_active_A} and  \ref{fig:RX_active_B} show numerical results for metasurfaces illuminated normally when the design reflection angles are $\theta_{\rm r}=30^\circ$  and $\theta_{\rm r}=70^\circ$, respectively. We can see how the amplitude of three reflected wave changes with the reflected angle according  to the theory. 
The surface impedance defined by  Eq.~(\ref{eq:rx_imp_active}) is a complex number [see Fig.~\ref{fig:RX_lossy_C}] whose real part takes positive (loss) and negative (gain) values. Obviously, the averaged over one period normal component of the total power is zero, meaning that the macroscopic system is  lossless.

Although active acoustic metamaterials have been studied in the literature \cite{review}, in general, the use of active and lossy elements is not desired in actual implementations for practical reasons. In order to simplify the design and implementation, the active-lossy behavior can be understood as a phenomenon of energy channeling, so it is not necessary to include active or lossy elements for implementing these metasurfaces, and a lossless implementation can be found. In order to overcome the fundamental deficiency of all conventional reflective metasurface and implement the required ``gain-loss'' response defined by Eq.~(\ref{eq:rx_imp_active}), the metasurface has to receive energy in the ``lossy'' regions, guide it along the surface, and radiate back in the ``active'' regions.  The energy channeling along  the metasurface corresponds to a non-local response.  In non-local metasurfaces the behavior of each element of the metasurface depends on the interaction with the neighbors,  so traditional techniques based on the individual design of each meta-atom cannot be used.
As it was demonstrated in \cite{last} for electromagnetic metasurfaces, properly designing the inhomogeneous impedance of a lossless metasurface it is possible to obtain the required non-local response. The operating principle is similar to leaky waves antennas \cite{Leaky1,Leaky2}, where periodical perturbations allow coupling between guided waves and propagating waves in free space. 

\begin{figure}[]
	\centering
	\begin{minipage}{1\columnwidth}
		\subfigure[]{\includegraphics[width=0.8\linewidth]{FIG10_C.pdf}\label{fig:RX_tubes_C}}
	\end{minipage}
	\begin{minipage}{1\columnwidth}
		
		\subfigure[]{\includegraphics[width=0.45\linewidth]{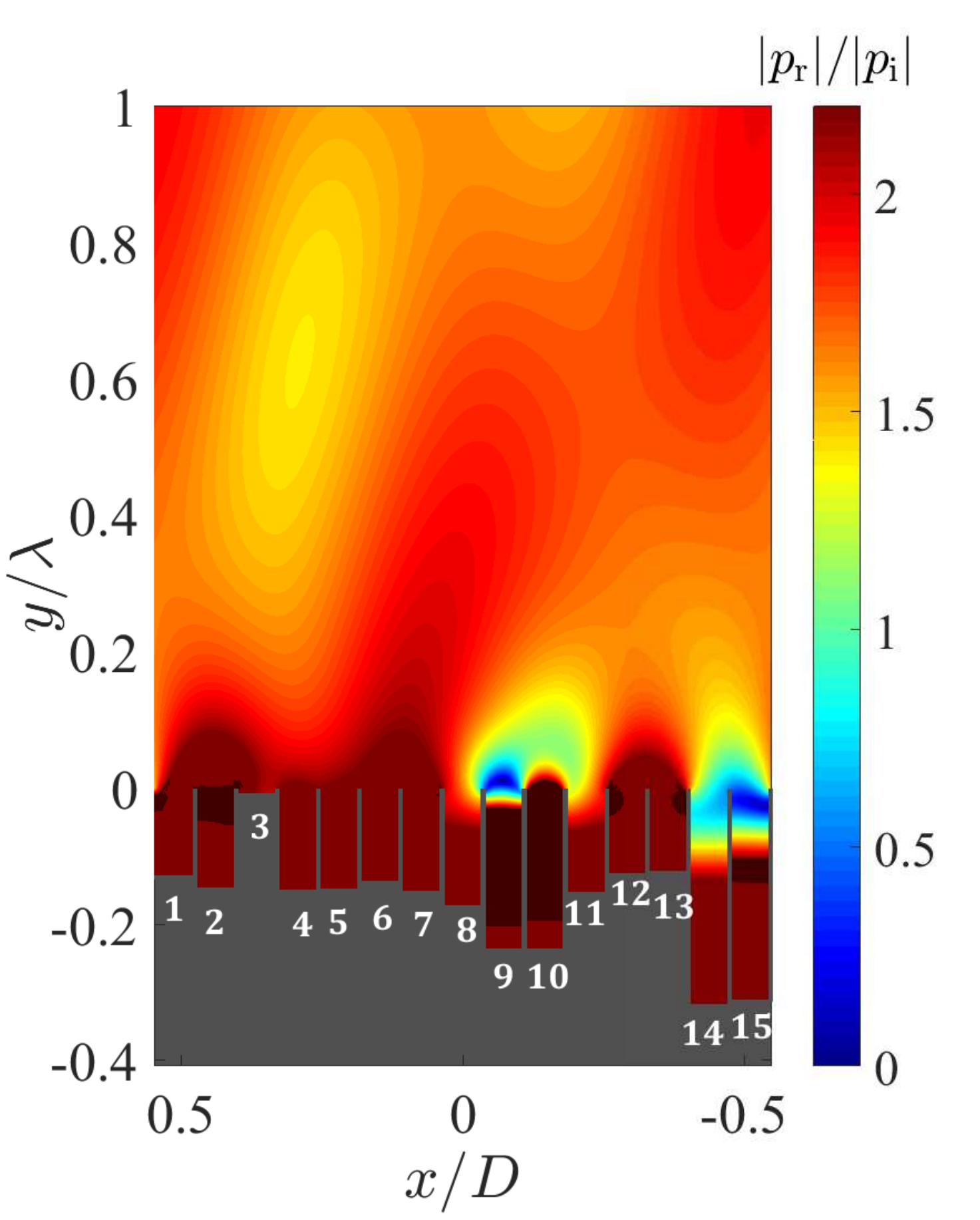}\label{fig:RX_tubes_B}}
		\subfigure[]{\includegraphics[width=0.5\linewidth]{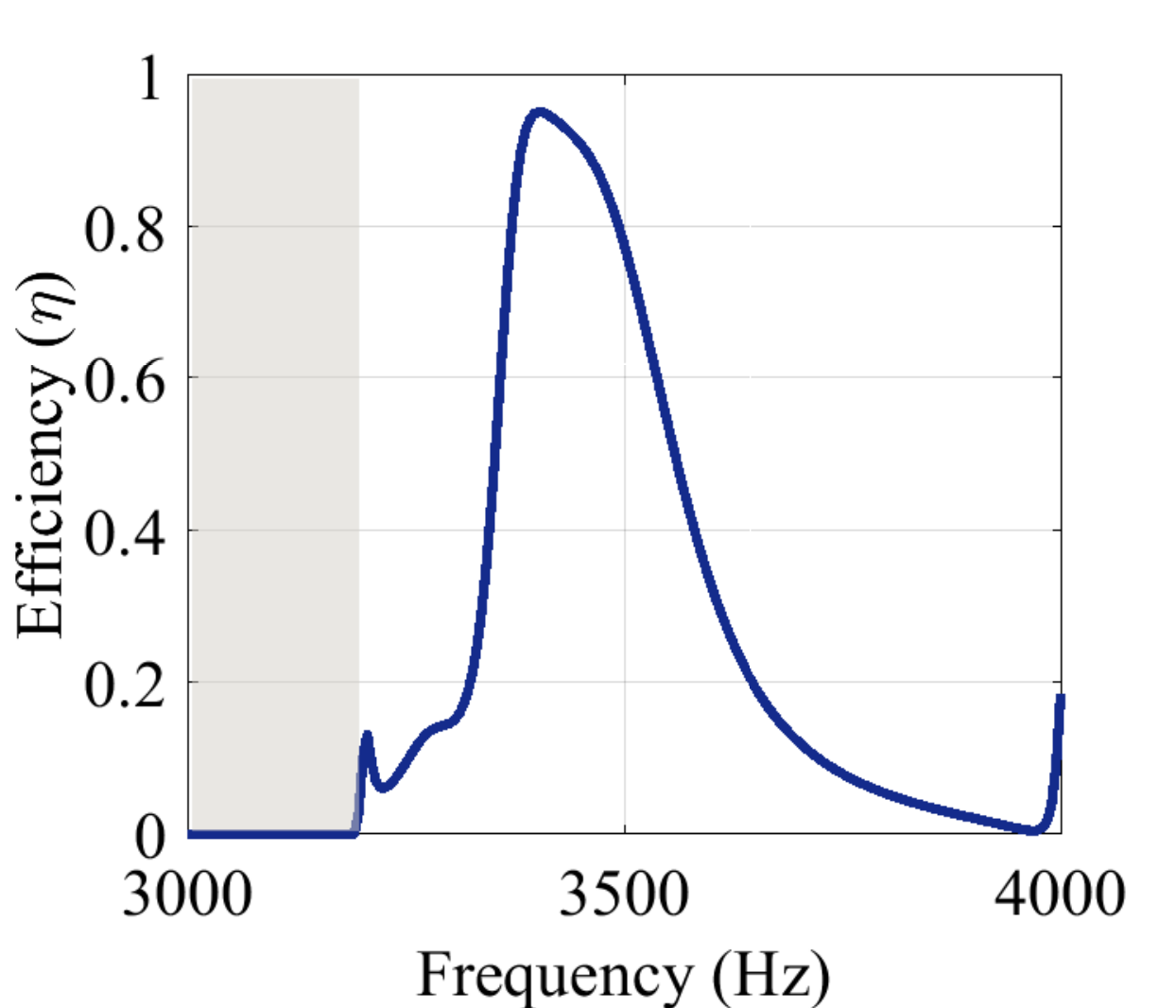}\label{fig:RX_tubes_D}}
	\end{minipage}
	\caption{  
		Non-local design of an anomalous reflector for  $\theta_{\rm i}=0^\circ$ and $\theta_{\rm r}=70^\circ$ with 95\% of efficiency: Real part (a) and magnitude (b) of the scattered field. 
		(c) Bandwidth analysis for an actual implementation of the anomalous reflector designed at 3400 Hz.  }\label{fig:RX_tubes}
\end{figure}

In what follows, we present a proof of concept of a non-local design with high efficient performance when  $\theta_{\rm i}=0^\circ$ and $\theta_{\rm r}=70^\circ$. For designing the non-local acoustic reflector, we use as a first approximation the imaginary part of the complex impedance described by Eq.~(\ref{eq:rx_imp_active}). This approach allows one to design an array of lossless elements by using conventional techniques which will produce a local phase shift according to $\Gamma=\frac{j\Im(Z_{\rm s})-Z_{\rm i}}{j\Im(Z_{\rm s})+Z_{\rm i}}$. Particularly, each element is implemented by a rigidly ended waveguide whose impedance can be calculated as
\begin{equation}
Z_{\rm stub}=-jZ_0 \cot(kl_n)
\end{equation}
where $l_n$ is the length of the {\it n}-th element. Once the length of each stub has been fixed (15 particles in the example presented in this work), we run a numerical optimization setting as a goal full reflection in the desired direction. The objective of this optimization is to tailor the coupling effects between particle, using the evanescent fields as a mechanism for channeling the energy.  After the optimization process the lengths of the stubs are  $0.1304\lambda$,  $ 0.1480  \lambda$,  $ 0.0103\lambda$,  $0.1519 \lambda$,  $0.1511\lambda$,  $0.1397    \lambda$,  $ 0.1541    \lambda$,  $0.1750   \lambda$,  $ 0.2389    \lambda$,  $0.2390\lambda$,  $ 0.1558\lambda$,  $0.1283   \lambda$,  $0.1252  \lambda$, $0.3203  \lambda$, and $0.3135\lambda$, respectively. The efficiency of the optimized metasurface is 95\%.  Figure \ref{fig:RX_tubes_C} shows the real part of the scattered field, where we can see the plane wave reflected in the desired direction.  Figure \ref{fig:RX_tubes_B} represents the magnitude of the scattered field. In the vicinity of the metasurface, strong evanescent fields are created, as is required for theoretically perfect performance.The corrugated surface  can be considered as a lossless interface able to receive power in some regions, guide it by surface waves excited in the metasurface and radiate the energy back into the desired direction.

For a more complete analysis of the non-local anomalous reflector, we study the efficiency as a function of frequency. In particular, we consider the previous design of the non-local metasurface  when the operation frequency is 3400~Hz. The results of this analysis are summarized in Fig.~\ref{fig:RX_tubes_D}.  We obtain the maximum efficiency at the design frequency. The efficiency is  defined as the power redirected into the first diffracted mode ($n=1$) which corresponds to $\theta_{\rm r}=70^\circ$  at the design frequency. The efficiency remains higher than 0.5 from 3350~Hz to 3550~Hz ($\approx 6 \%$ of the operating frequency).

\section{Acoustic metasurfaces for anomalous transmission}

\subsection{Design based on the generalized Snell's law}

\begin{figure}[]
	\centering
	      \subfigure[]{\includegraphics[width=0.45\linewidth]{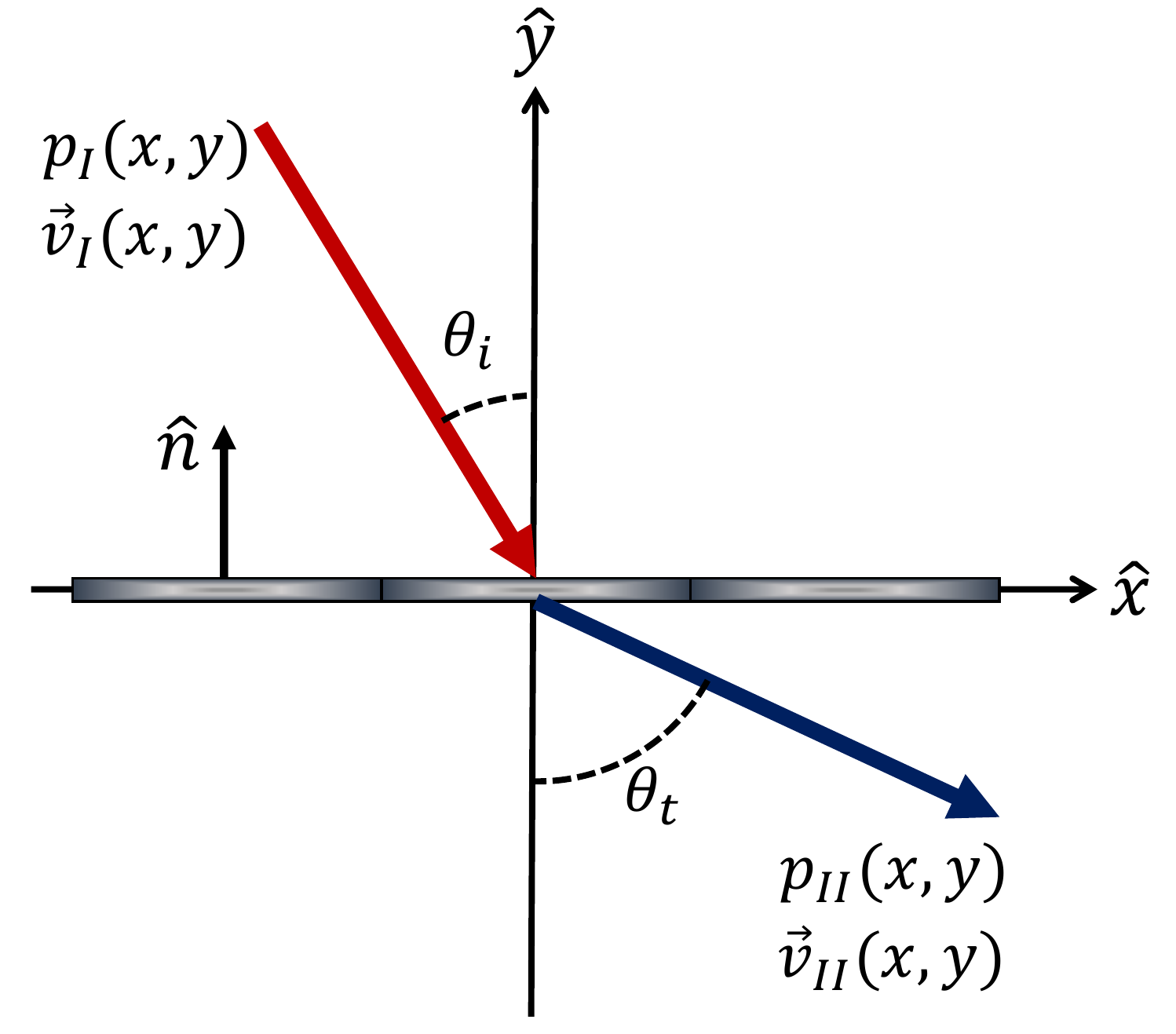}\label{fig:FIG5_A}}
	\raisebox{0.5\height}{\subfigure[]{\includegraphics[width=0.45\linewidth]{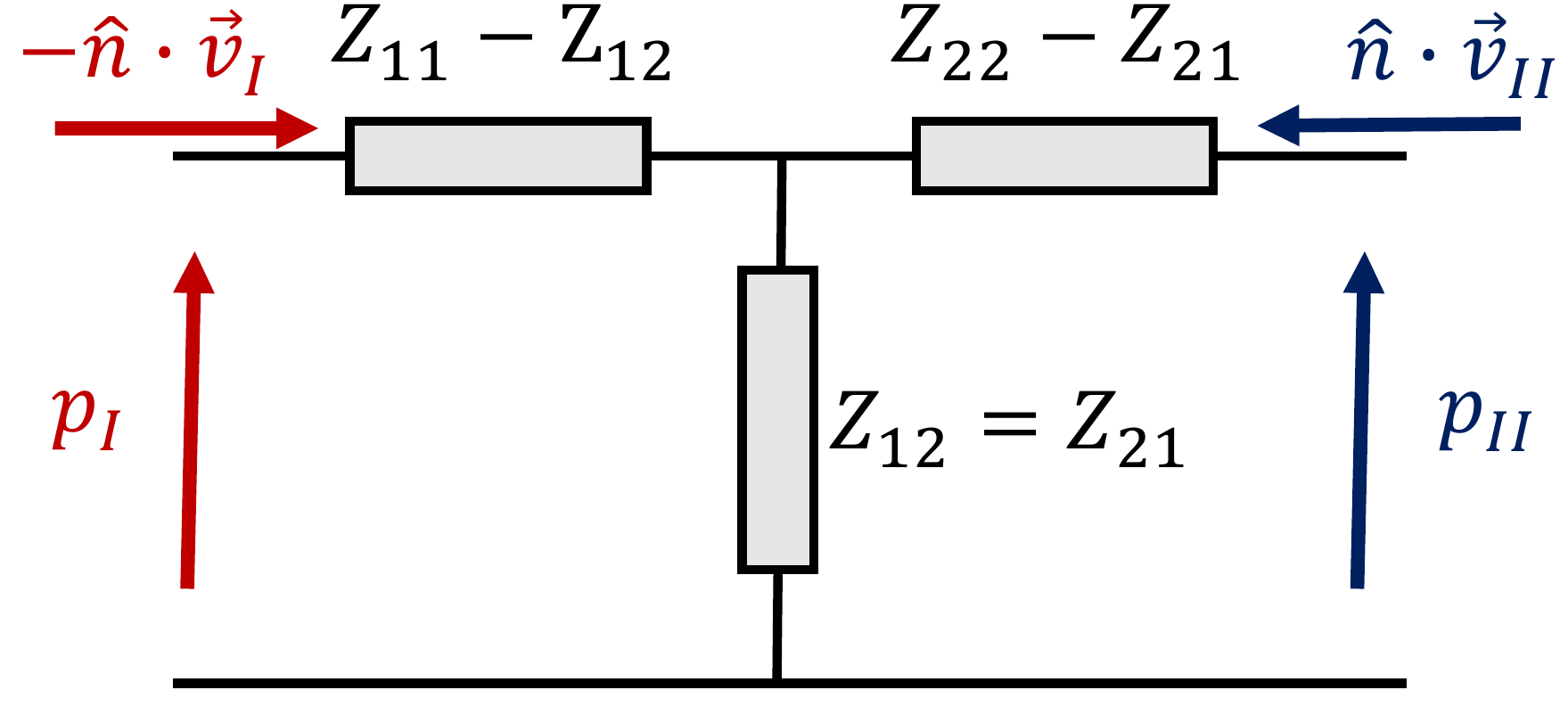}\label{fig:FIG5_B}}}
	\caption{(a) Schematic representations of the metasurface behavior for the anomalous transmission scenario. (b) Equivalent circuit for refractive metasurfaces.}\label{fig:FIG5}
\end{figure}

As in the above study of reflective metasurfaces, we start with the analysis of the conventional refractive gradient index metasurfaces  based on the generalized Snell's law. If we consider the scenario illustrated in Fig.~\ref{fig:FIG5_A}, the pressure field above and beyond the metasurface can be written as
\begin{eqnarray}p_{\rm I}(x,y)=p_0e^{-jk \sin{\theta_{\rm i}}x}e^{jk \cos{\theta_{\rm i}}y},\\
 p_{\rm II}(x,y)=Ap_0e^{-jk \sin{\theta_{\rm t}}x}e^{jk \cos{\theta_{\rm t}}y}, \end{eqnarray}
where $p_0$ is the amplitude of the incident plane wave, $\theta_{\rm i}$ and $\theta_{\rm t}$ are the incidence and transmission angles, respectively, $k=\omega/c$ is the wavenumber at the operation frequency, and $A$ is the coefficient which relates the amplitudes of the incident and transmitted waves. 
The velocity vectors at both sides of the metasurface can be expressed as
\begin{eqnarray}
\vec{v}_{\rm I}(x,y)=\frac{p_{\rm I}(x,y)}{Z_0}\left(\sin{\theta_{\rm i}}\hat{x}-
\cos{\theta_{\rm i}}\hat{y}\right)\\
\vec{v}_{\rm II}(x,y)=\frac{ p_{\rm II}(x,y)}{Z_0}\left(\sin{\theta_{\rm t}}\hat{x}-\cos{\theta_{\rm t}}\hat{y}\right)
\end{eqnarray}
with $Z_0=c\rho$ being the characteristic impedance of the background medium. 
Pressure and velocity at both side of the metasurface can be related by using the specific impedance matrix as
\begin{equation}
\begin{bmatrix}
p_{\rm I}(x,0)    \\
p_{\rm II}(x,0)
\end{bmatrix}=
\begin{bmatrix}
Z_{11}     & Z_{12}  \\
Z_{21}     & Z_{22}
\end{bmatrix}\begin{bmatrix}
-\hat{n} \cdot\vec{v}_{\rm I}(x,0)    \\
 \hat{n} \cdot\vec{v}_{\rm II}(x,0)
\end{bmatrix}.\label{eq:Tx_BC}
\end{equation}
In the most general linear case and assuming reciprocity ($Z_{12}=Z_{21}$), the relation between the acoustic field at both side of the metasurfaces can be modeled by the equivalent circuit represented in Fig.~\ref{fig:FIG5_B}. 
Conventional refractive metasurfaces are designed in such a way that each meta-atom introduces a local phase-shift in transmission according to 
\begin{equation}
t=\frac{p_{\rm II}(x,0)}{p_{\rm I}(x,0)}=e^{j\Phi_x},\label{eq:Tx_gradient}
\end{equation}
where $\Phi_x=k\sin{\theta_{\rm i}}x-k\sin{\theta_{\rm t}}x$ is the linearly-varying phase of the local transmission coefficient $t$. In the known designs symmetric meta-atoms ($Z_{11}=Z_{22}$) are used for the implementation of the desired transmission coefficient at every point of the metasurface. The relation between the $Z$-matrix elements and the local transmission coefficient can be expressed as \cite{Pozar}  
\begin{eqnarray}
Z_{11}=Z_{22}=Z_{\rm i} \frac{1+t^2}{1-t^2}=j\frac{Z_0}{\cos{\theta_{\rm i}}}\cot(\Phi_x),\label{eq:Tx_conventional2}\\
Z_{12}=Z_{\rm i}\frac{2t}{1-t^2}=j\frac{Z_0}{\cos{\theta_{\rm i}}}\frac{1}{\sin(\Phi_x)}.\label{eq:Tx_conventional1}
\end{eqnarray}
These impedances define the behavior of conventional designs. We can see  that the impedances are purely imaginary, meaning that lossless implementations are possible \cite{Tx1,Tx2,Tx3,Tx4,Tx5}. 

\begin{figure}[h!]
	\centering
	\subfigure[]{\includegraphics[width=0.65\linewidth]{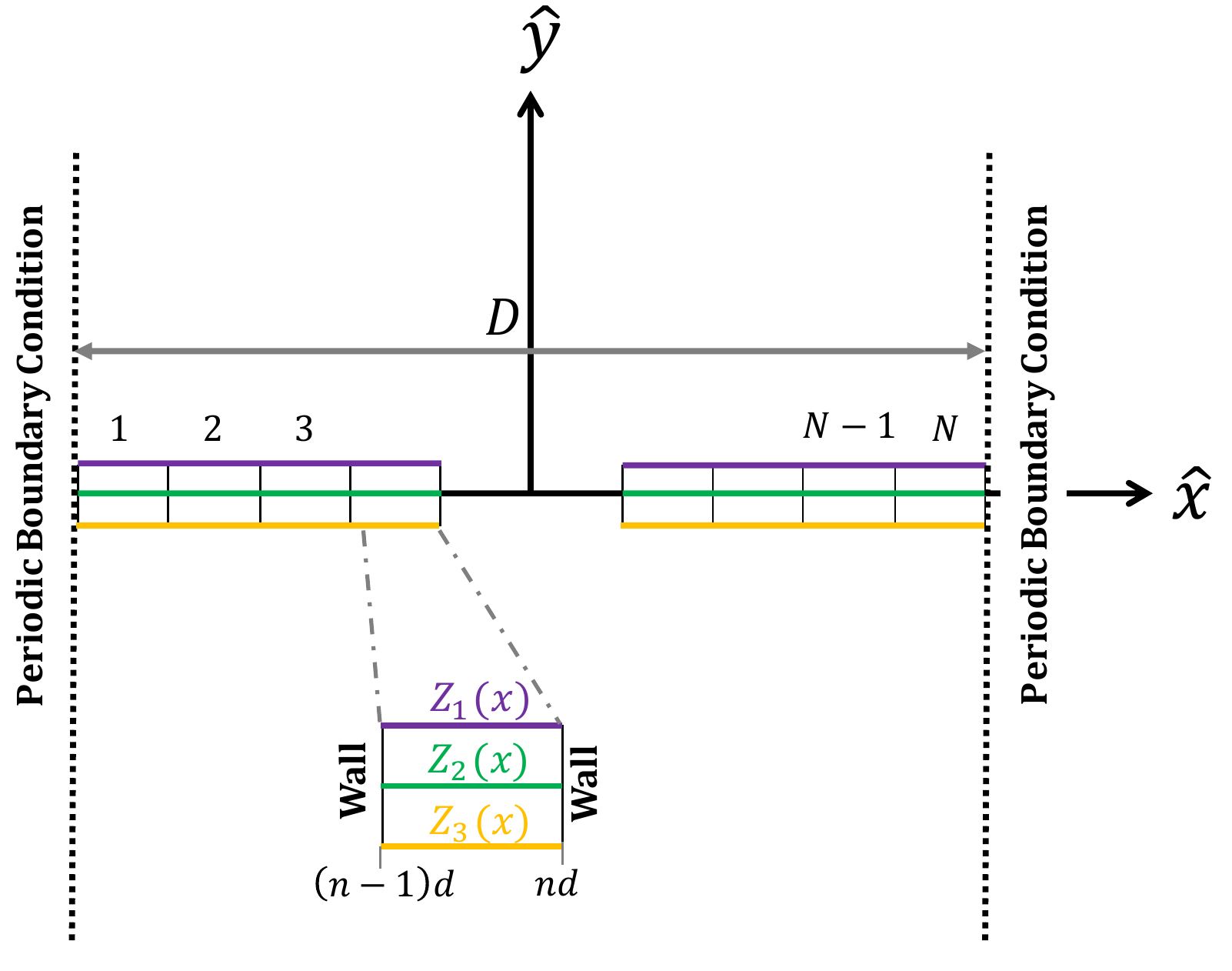}\label{fig:FIG6_A}}
	\subfigure[]{\includegraphics[width=0.3\linewidth]{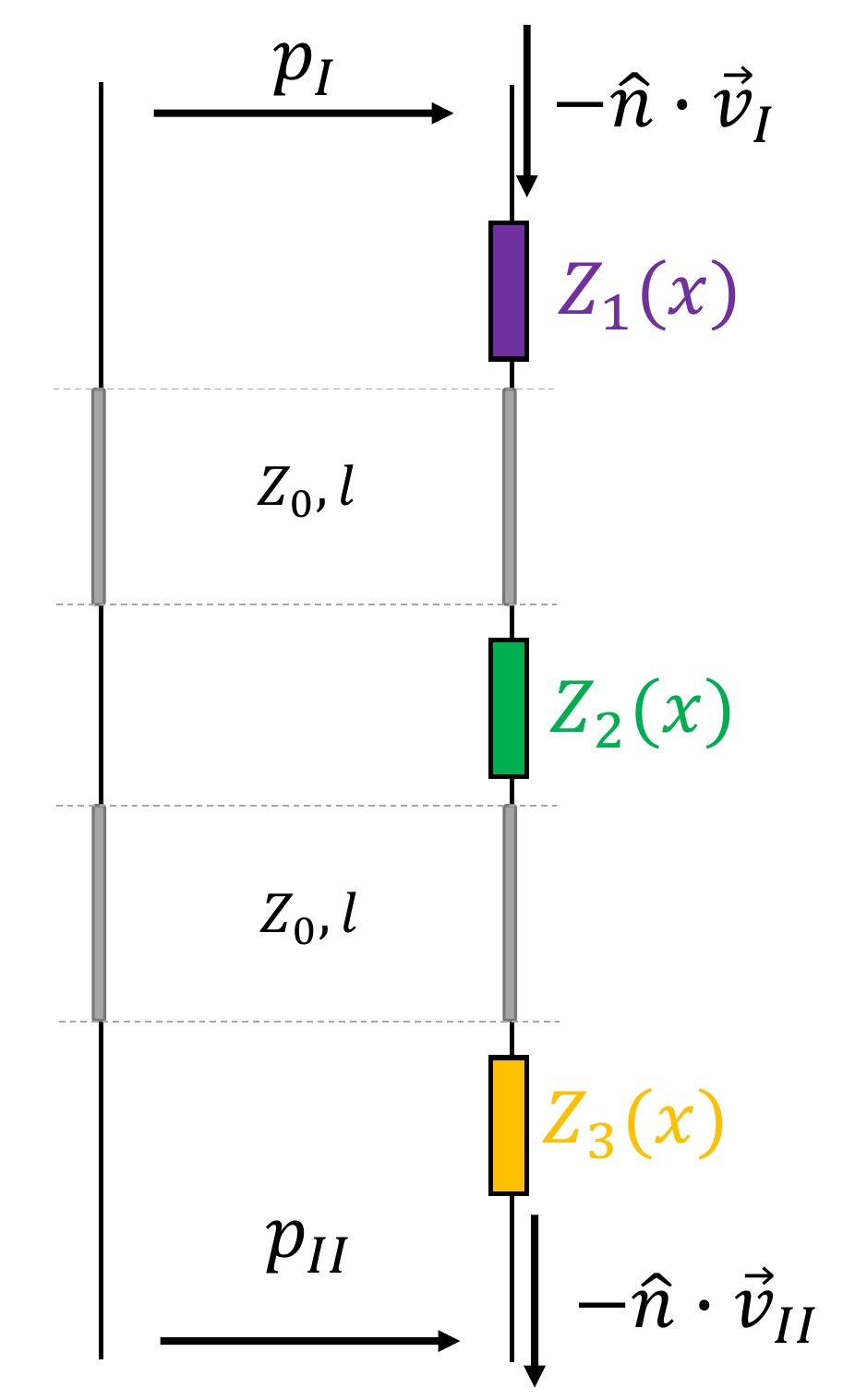}\label{fig:FIG6_B}}
	\caption{(a) Schematic representations of the metasurface topology (b) Equivalent circuit for the proposed meta-atoms.}\label{fig:FIG6}
\end{figure}

In this work and as a proof of concept we propose a simple implementation of the meta-atoms based on clamped rectangular membranes. Each membrane can be modeled as a series {\it LC}-resonator controlled by its acoustic mass and compliance \cite{Leaky1, Leaky2}. Particularly, each meta-atom  consists of three membranes separated by distance $l$ [see Fig.~\ref{fig:FIG6}]. The period of the metasurface is divided into $N$ unit-cells (the width of the meta-atoms is $D/N$) and rigid walls are introduced between the meta-atoms to avoid coupling between them and prevent excitation of guided modes between membranes. By independently tuning the response of each membrane we can obtain the desired response of the meta-atoms.

The equivalent circuit for the proposed implementation is shown in  Fig.~\ref{fig:FIG6_B}, where the membranes are modeled as reactive impedances in series. For designing the membranes we need to know the relation between their sheet impedances and the $Z$-matrix.  The response of a meta-atom can be expressed in terms of the transmission matrices of membranes and empty spacings between them:
\begin{equation}
\begin{bmatrix}
p_{\rm I}(x,0)    \\
-\hat{n} \cdot\vec{v}_{\rm I}(x,0)
\end{bmatrix}=
\begin{bmatrix}
A     & B  \\
C     & D
\end{bmatrix}\begin{bmatrix}
p_{\rm II}(x,0)    \\
-\hat{n} \cdot\vec{v}_{\rm II}(x,0)
\end{bmatrix},
\end{equation}
where
\begin{equation}
\left[ {\begin{array}{cc}
	A & B \\
	C & D      \end{array} } \right]=M_{Z1}M_{T}M_{Z2}M_{T}M_{Z3}
\end{equation}
with 
\begin{equation}
M_{Zi}=\left[ {\begin{array}{cc}
	1 & Z_i \\
	0 & 1      \end{array} } \right], \qquad i=1,2,3
\end{equation}
and
\begin{equation}
M_{T}=\left[ {\begin{array}{cc}
	\cos(kl) & jZ_0\sin(kl) \\
	j\frac{1}{Z_0}\sin(kl) & \cos(kl)      \end{array} } \right].
\end{equation}
The three elements of the ABCD-matrix needed for the definition of meta-atoms are
\begin{eqnarray}
\begin{split}
A=	\cos^2(kl)-	\sin^2(kl)\left(1+\frac{Z_2Z_1}{Z_0^2}\right)\\+j	\cos(kl)\sin(kl)\left(\frac{2Z_1+Z_2}{Z_0}\right)\end{split}\\
C=j2Y_0	\cos(kl)\sin(kl)-\frac{Z_2}{Z_0^2}\sin^2(kl)\\
\begin{split}D=\cos^2(kl)-\frac{Z_3Z_2}{Z_0^2}\sin^2(kl)\\+2j\frac{Z_3+Z_2}{Z_0}\cos(kl)\sin(kl).\end{split}
\end{eqnarray}
On the other hand, we can write the desired response modeled by the $Z$-matrix in terms of the ABCD-matrix \cite{Pozar} as
\begin{equation}
\left[\begin{array}{cc}
A&B\\
C&D\\
\end{array}\right]=
\left[\begin{array}{cc}
\frac{Z_{11}}{Z_{21}}&\frac{|Z|}{Z_{12}}\\
\frac{1}{Z_{12}}&\frac{Z_{22}}{Z_{12}}\\
\end{array}\right]\label{eqt:ABCD&Z}
\end{equation}
with $|Z|=Z_{11}Z_{22}-Z_{12}^2$.
Finally, equating each element of both ABCD-matrices, we can obtain the relations which define the membranes:
\begin{eqnarray}
Z_{1}(x)=Z_{11}+Z_{12}+jZ_0\cot(kl),\\
Z_{2}(x)=j2Z_0\cot(kl)-\frac{Z_0^2}{Z_{12}}\frac{1}{\sin^2(kl)},\\
Z_{3}(x)=Z_{22}+Z_{12}+jZ_0\cot(kl).
\end{eqnarray}

\begin{figure}[h!]
	\centering
	\begin{minipage}{0.7\columnwidth}
		\subfigure[]{\includegraphics[width=\linewidth]{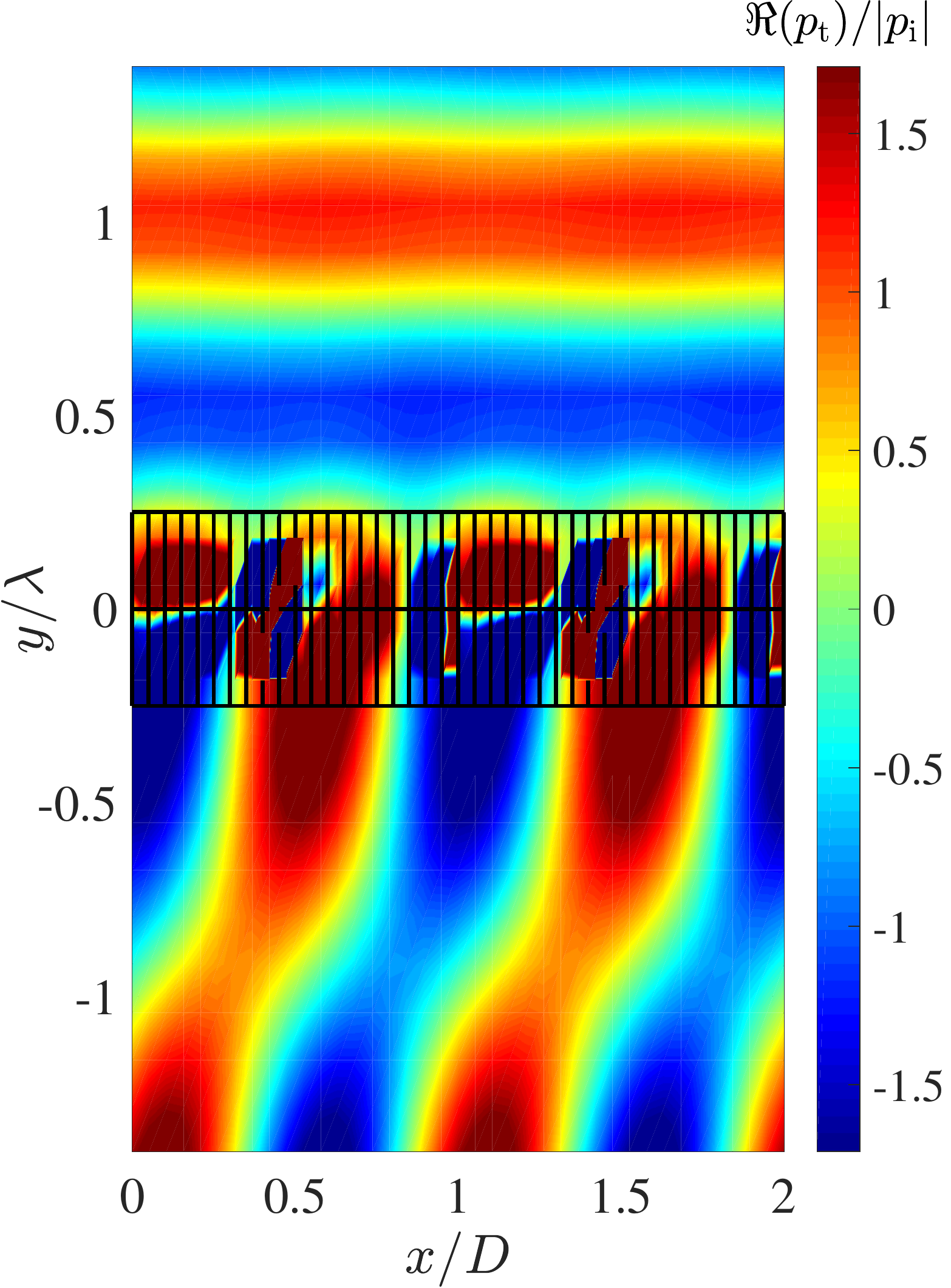}\label{fig:Tx_con_A}}
	\end{minipage}
	\begin{minipage}{1\columnwidth}
		\subfigure[]{\includegraphics[width=0.46\linewidth]{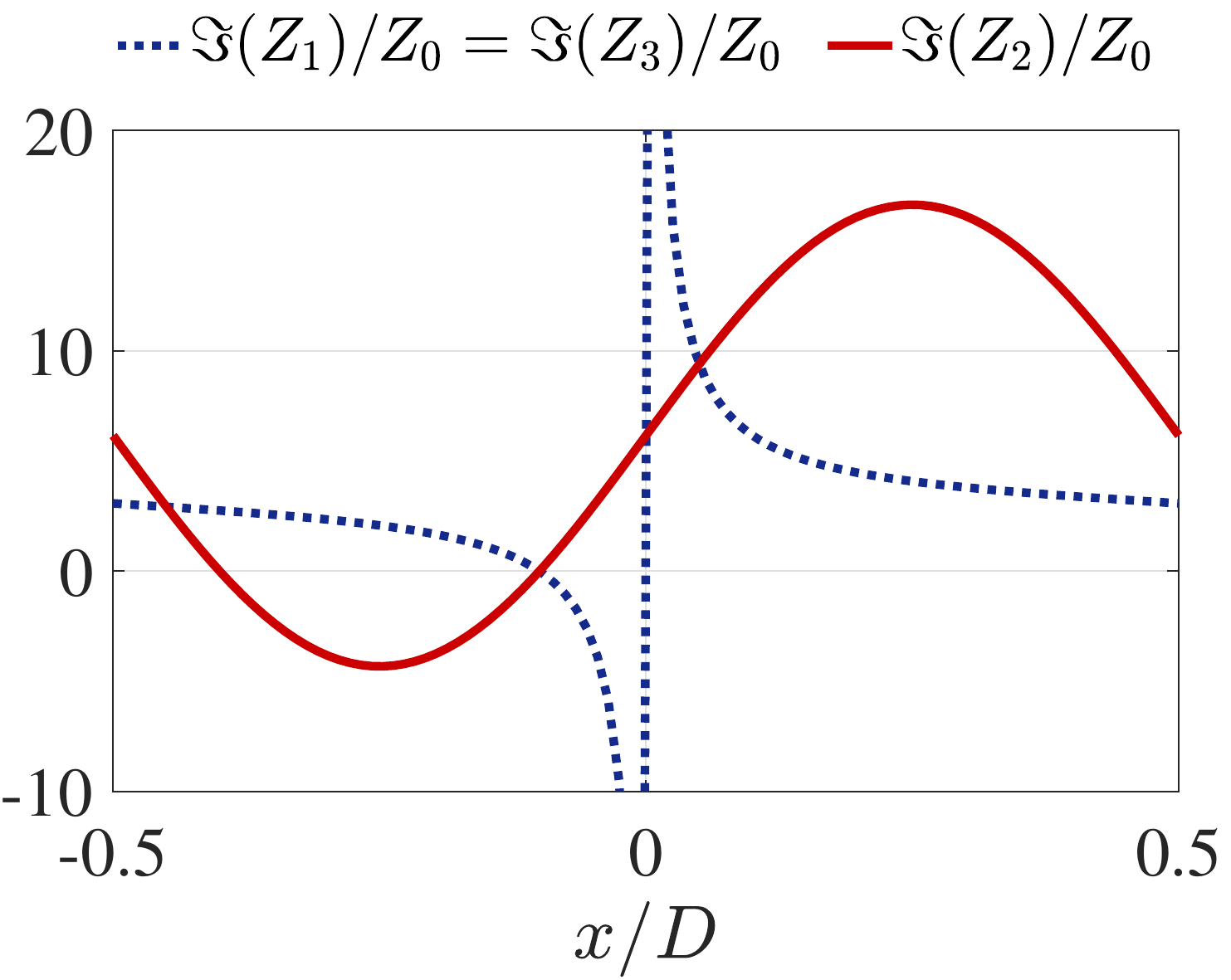}\label{fig:Tx_con_C}}
		\subfigure[]{\includegraphics[width=0.46\linewidth]{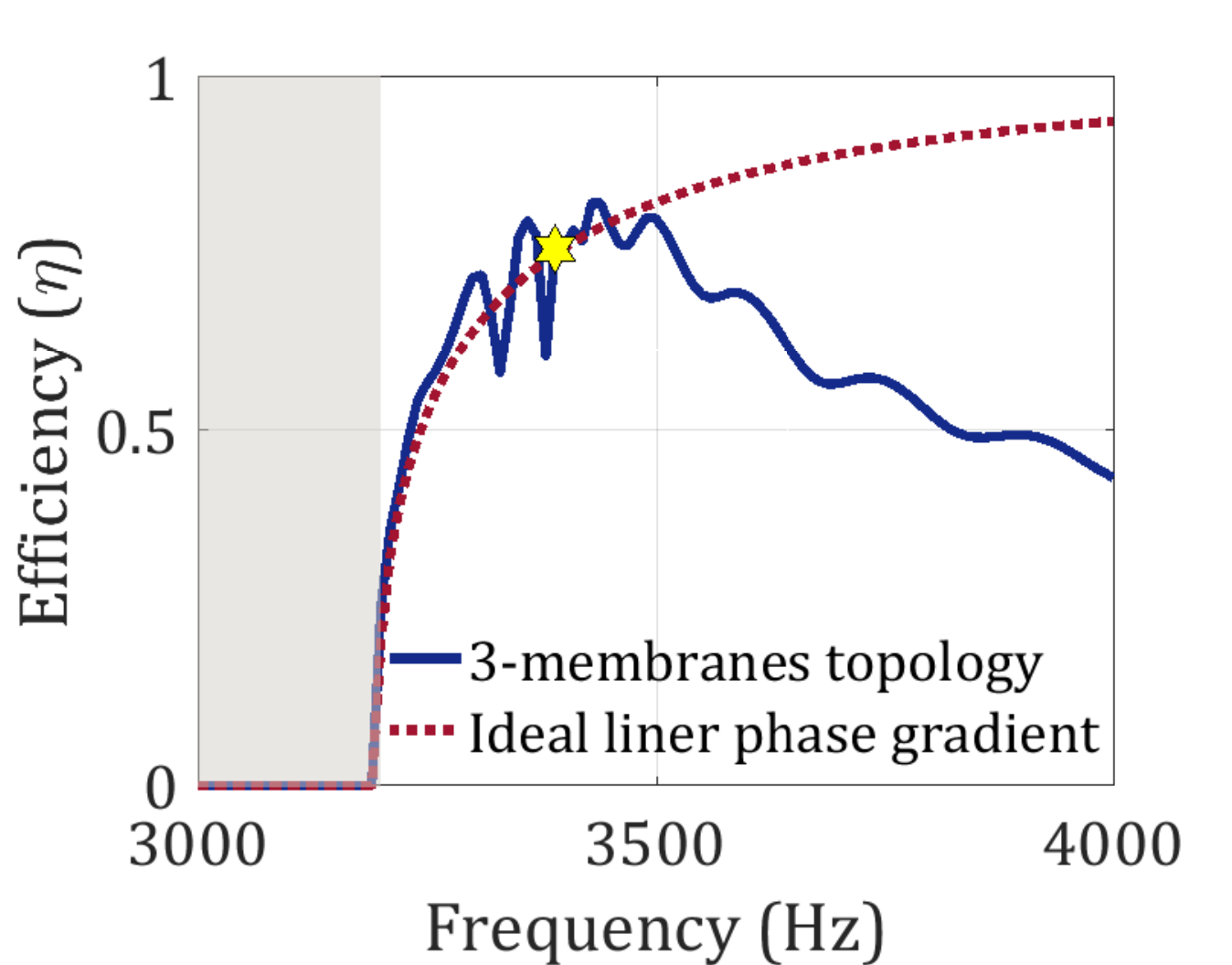}\label{fig:Tx_con_D}}
	\end{minipage}
	\caption{
		Refractive metasurface based on a linear phase gradient:}(a) Real part of the total pressure field for a conventional metasurface when  $\theta_{\rm i}=0^\circ$ and $\theta_{\rm t}=70^\circ$; (b) Impedance which model the three membranes of the meta-atoms; (c) Bandwidth of the metasurface designed for 3400 Hz.\label{fig:Tx_con}
\end{figure}

Figure \ref{fig:Tx_con} shows the results of the numerical analysis for an acoustic metasurface based on the generalized Snell's law. Particularly, the study has been done for  $\theta_{\rm i}=0^\circ$, $\theta_{\rm t}=70^\circ$, and $l=\lambda/4$ assuming 20 elements per period. The impedances of the membranes are represented in Fig.~\ref{fig:Tx_con_C}. The response of the metasurface for normal illumination is illustrated in  Fig.~\ref{fig:Tx_con_A} where we can see that the wavefront of the transmitted wave is not perfect  and some perturbations due to parasitic waves appear.

For comparison, it is interesting to analyse also conventional refractive metasurfaces based on  a linear phase gradient as a function of the frequency. Let us first consider a refractive metasurface made of non-dispersive elements designed for implementing a linear phase gradient according to Eq.~(\ref{eq:Tx_gradient}). The theoretical response of the designed structure as a function of the frequency can be calculated by considering the change in the impedance due to the change in the direction of the diffracted mode for different frequencies $\theta_{\rm r}(f)=\arcsin{\frac{k\sin\theta_{\rm i}+2\pi/D}{k}}$. Following the same approach as for reflective metasurfaces we can calculate the amplitude of the transmitted wave by analyzing the impedance mismatch between the incident and transmitted waves, $T(f)=\frac{2\cos\theta_{\rm i}}{\cos\theta_{\rm i}+\cos\theta_{\rm r}(f)}$.  Consequently, the efficiency of the metasurface defined as the percentage of power sent into the desired diffracted mode can be calculated as $\eta(f)=T(f)^2\frac{\cos\theta_{\rm r}(f)}{\cos\theta_{\rm i}}$.  This efficiency is represented  in  Fig.~\ref{fig:Tx_con_D} with the red line. In actual implementations, the dispersion of the elements has to be taken into account in the analysis of the bandwidth.  Figure~\ref{fig:Tx_con_D} shows the results of a numerical study of the bandwidth for a metasurface designed for operation at 3400~Hz (yellow symbol) according to   Eqs.~(\ref{eq:Tx_conventional2}) and (\ref{eq:Tx_conventional1}) and implemented with the 3-membranes topology described above. We can clearly see that the theoretical performance of the metasurface is perturbed by the dispersive behavior of the constituent elements reducing the bandwidth of the designed metasurface.

\subsection{Asymmetric acoustic metasurfaces for perfect anomalous refraction}

\begin{figure}[h!]
	\centering
	\begin{minipage}{0.7\columnwidth}
		\subfigure[]{\includegraphics[width=1\linewidth]{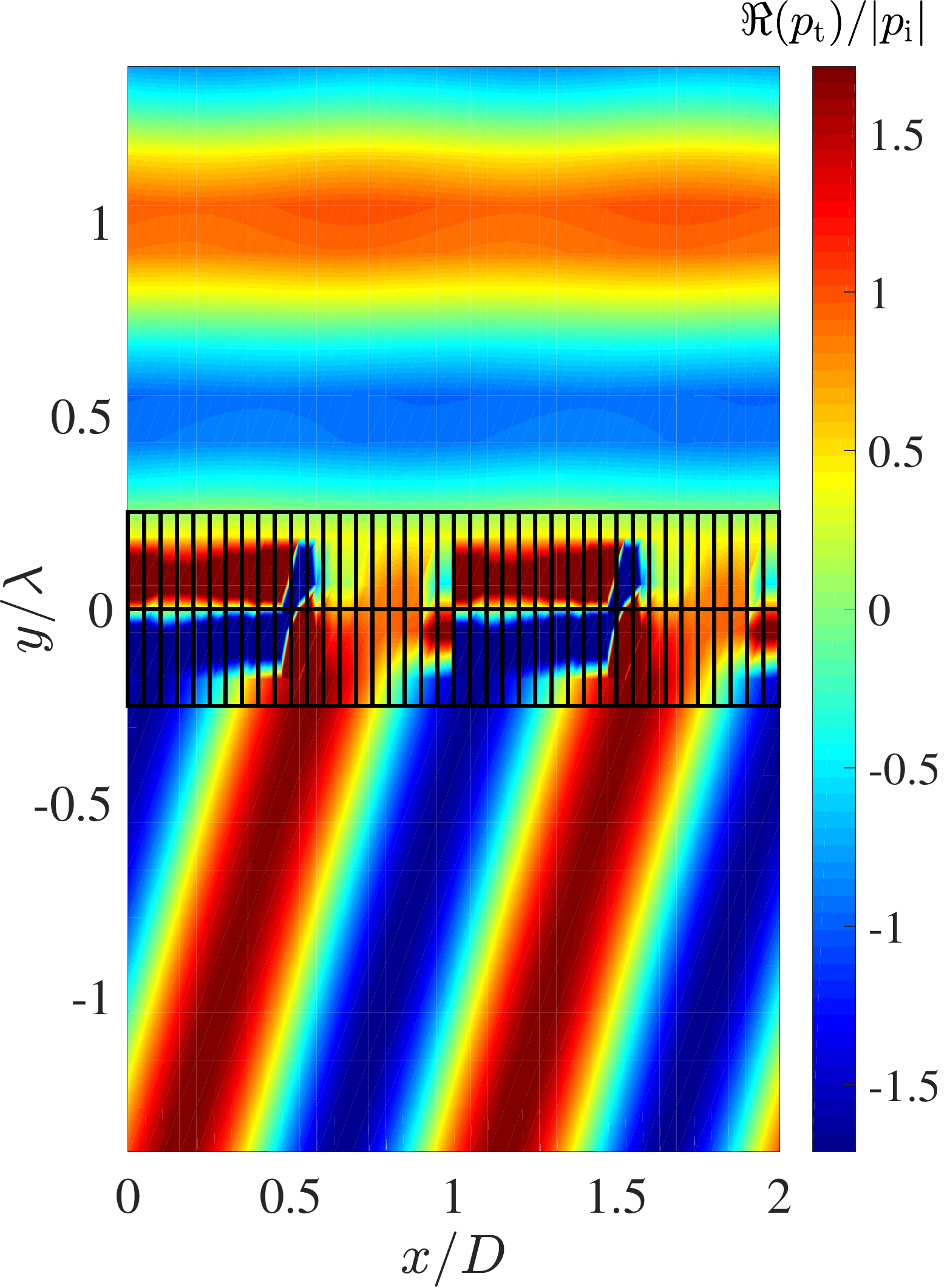}\label{fig:Tx_perfect_A}}
	\end{minipage}
	\begin{minipage}{1\columnwidth}
		\subfigure[]{\includegraphics[width=0.47\linewidth]{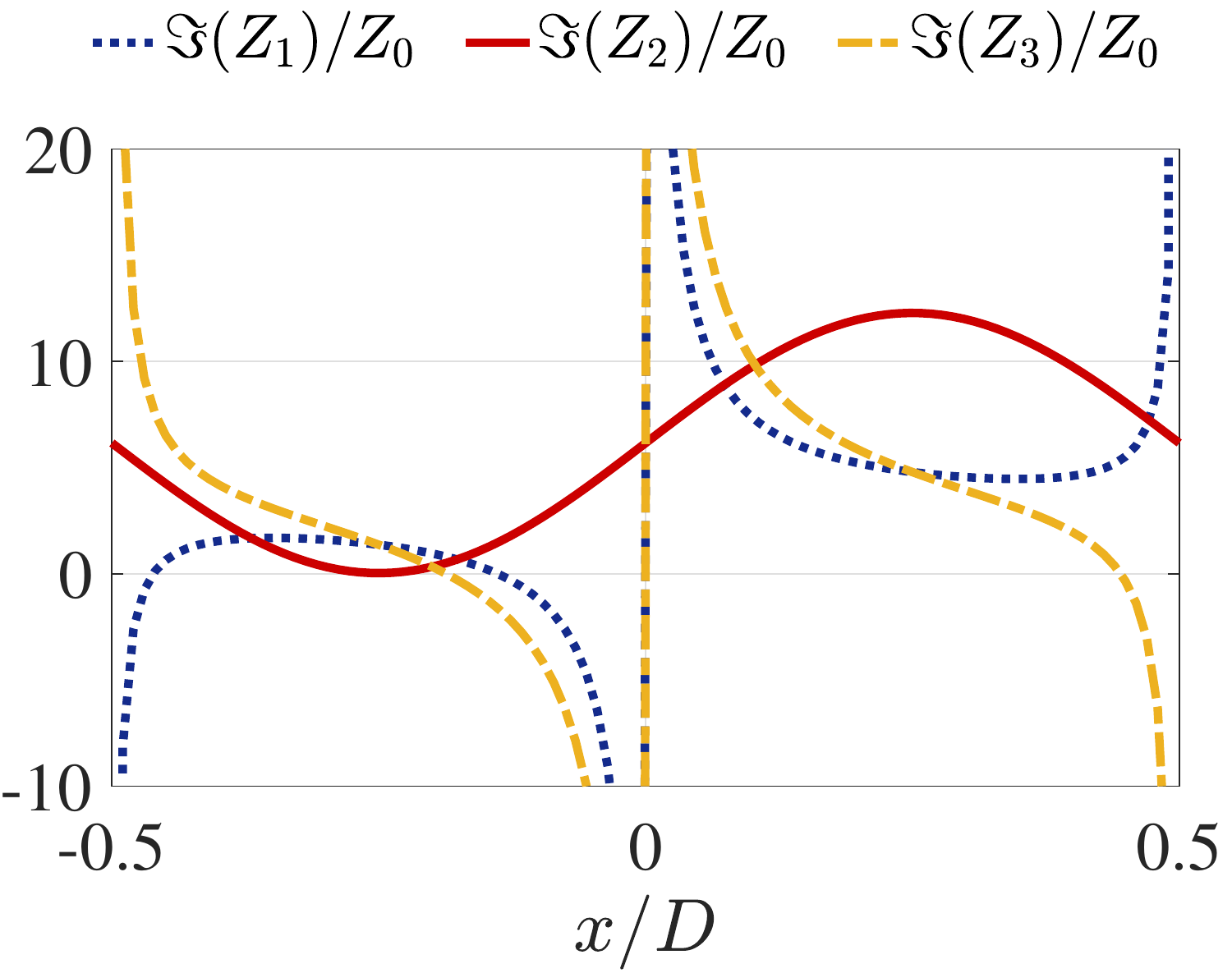}\label{fig:Tx_perfect_B}}
	   \subfigure[]{\includegraphics[width=0.47\linewidth]{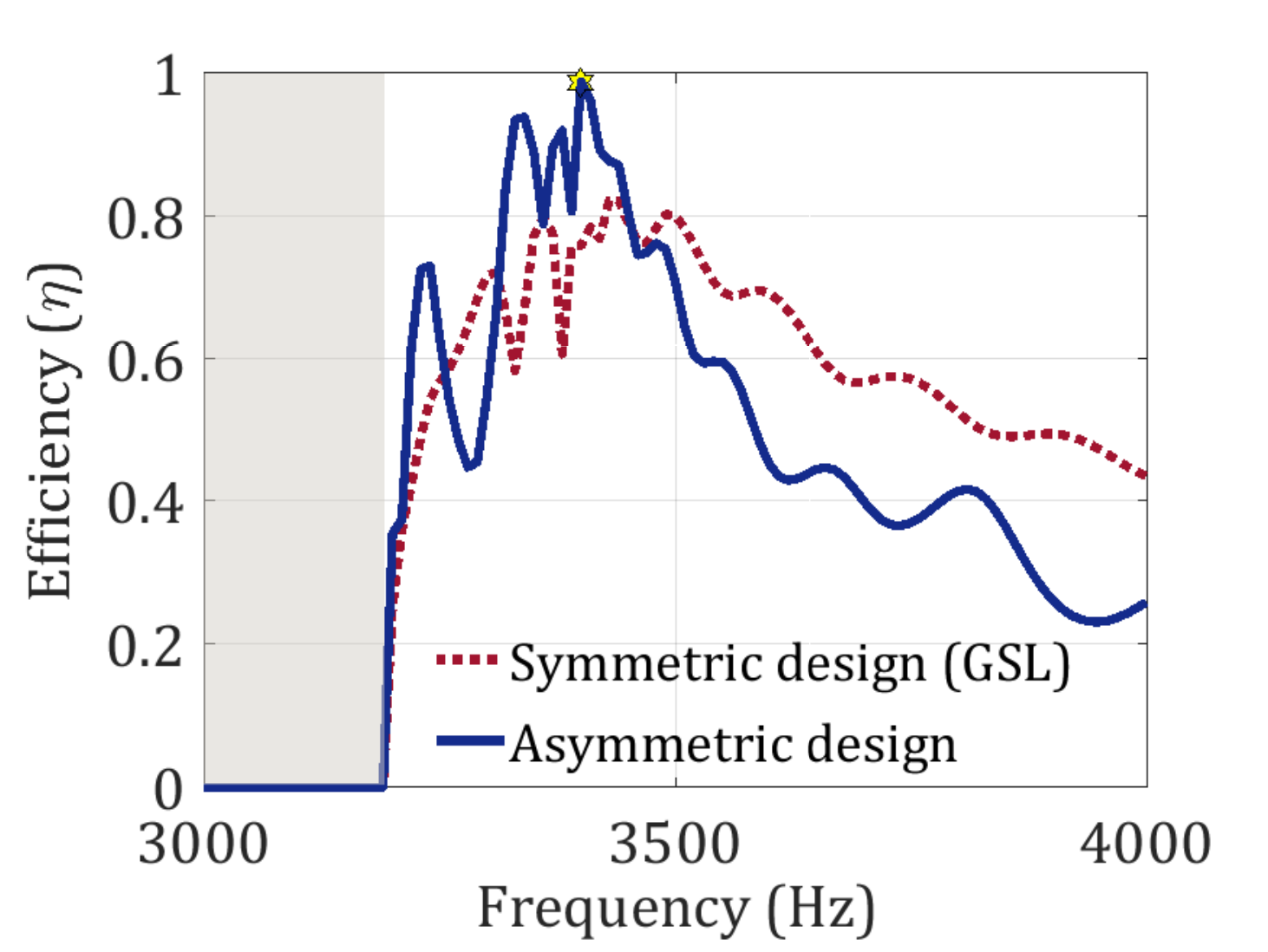}\label{fig:Tx_perfect_C}}
	\end{minipage}
	\caption{Asymmetric refractive metasurface. (a) Real part of the total pressure field for a perfect metasurface when  $\theta_{\rm i}=0^\circ$ and $\theta_{\rm t}=70^\circ$; (b) Impedance which model the three membranes of the meta-atoms; (c) Bandwidth of the metasurface designed for 3400 Hz. }\label{fig:Tx_perfect}
\end{figure}

As it was explained for reflective metasurfaces, for a perfect performance of the refractive metasurface we have to ensure that all the energy of the incident plane wave is carried away by the transmitted plane wave propagating in the desired direction.  This condition gives us the following relation:
\begin{equation}
\frac{p_0^2}{Z_0}\cos{\theta_{\rm i}}=A^2\frac{p_0^2}{Z_0}\cos{\theta_{\rm t}}.
\end{equation}
The amplitudes of the incident and transmitted waves have to be different, and the relation between them reads
\begin{equation}
A=\sqrt{\frac{\cos{\theta_{\rm i}}}{\cos{\theta_{\rm t}}}}. \label{eq:Tx_amplitudes}
\end{equation}
If we impose this amplitude for the transmitted wave and look for  a lossless reciprocal solution  ($Z_{11}=jX_{11}$, $Z_{22}=jX_{22}$, and $Z_{21}=Z_{12}=jX_{21}$),  the boundary conditions given by Eq.~(\ref{eq:Tx_BC}) simplify to
\begin{eqnarray}
1=jX_{11}\frac{\cos{\theta_{\rm i}}}{Z_0}- j X_{21} A\frac{\cos{\theta_{\rm r}}}{Z_0}e^{j\Phi_x},\\
Ae^{j\Phi(x)}=jX_{21}\frac{\cos{\theta_{\rm i}}}{Z_0}- j X_{22} A\frac{\cos{\theta_{\rm r}}}{Z_0}e^{j\Phi_x}.
\end{eqnarray}
Separating the real and imaginary parts, can write:
\begin{eqnarray}
1= X_{21} A\frac{\cos{\theta_{\rm r}}}{Z_0}\sin{(\Phi_x)},\\
0=X_{11}\frac{\cos{\theta_{\rm i}}}{Z_0}-  X_{21} A\frac{\cos{\theta_{\rm r}}}{Z_0}\cos{(\Phi_x)},\\
A\cos{\Phi(x)}= X_{22} A\frac{\cos{\theta_{\rm r}}}{Z_0}\sin{(\Phi_x)},\\
A\sin{\Phi(x)}=X_{21}\frac{\cos{\theta_{\rm i}}}{Z_0}- X_{22} A\frac{\cos{\theta_{\rm r}}}{Z_0}\cos{(\Phi_x)}.
\end{eqnarray}
Solving this system of equations, we finally find the elements of the corresponding $Z$-matrix: 
\begin{eqnarray}
Z_{11}=j\frac{Z_0}{\cos{\theta_{\rm i}}}\cot{(\Phi_x)},\label{eq:Txperfec1}\\
Z_{22}=j\frac{Z_0}{\cos{\theta_{\rm t}}}\cot{(\Phi_x)},\label{eq:Txperfec2}\\
Z_{12}=j\frac{Z_0}{\sqrt{\cos{\theta_{\rm i}}\cos{\theta_{\rm t}}}}\frac{1}{\sin{(\Phi_x)}}\label{eq:Txperfec3}.
\end{eqnarray}
The first difference as compared with the conventional design is that the meta-atoms are not symmetric ($Z_{11}\ne Z_{22}$).  This asymmetric response is comparable with the bianisotropic requirements described for the electromagnetic counterpart \cite{synthesis}. The same three-membranes topology can be used for the implementation of this new design. However, as we can see from  Fig.~\ref{fig:Tx_perfect_A}-\ref{fig:Tx_perfect_B}, $Z_1\ne Z_3$ in order to realize the required asymmetry. Figure~\ref{fig:Tx_perfect_A} shows the results of numerical simulations when  $\theta_{\rm i}=0^\circ$, $\theta_{\rm t}=70^\circ$. Clearly, the proposed design generates a perfect plane wave in the desired direction. 

Figure~\ref{fig:Tx_perfect_C} shows the results of a numerical study of the bandwidth and the comparison between a symmetric design [according to Eqs.~(\ref{eq:Tx_conventional2}) and (\ref{eq:Tx_conventional1})] and an asymmetric design [according to Eqs.~(\ref{eq:Txperfec1}), (\ref{eq:Txperfec2}), and (\ref{eq:Txperfec3})]. The operational frequency for both designs is 3400 Hz. The efficiency of the asymmetric design is higher than of the  GSL-based design in the frequency range between 3315~Hz and 3470~Hz; then the efficiency of the asymmetric design decrease faster.   The efficiency of the  proposed design is higher than 0.5 in the frequency range between 3295~Hz and 3585~Hz (8.5 \% of the operating frequency).

\begin{figure}[h!]
\centering
	\begin{minipage}{0.6\columnwidth}
	\subfigure[]{\includegraphics[width=0.45\linewidth]{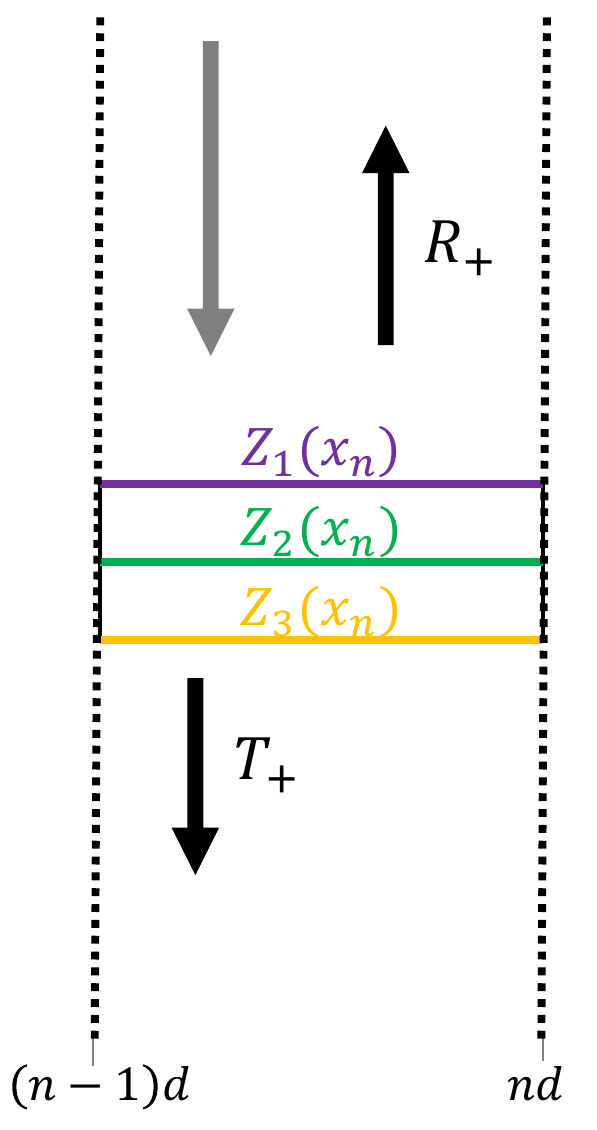}\label{fig:FIG9A}}
	\subfigure[]{\includegraphics[width=0.45\linewidth]{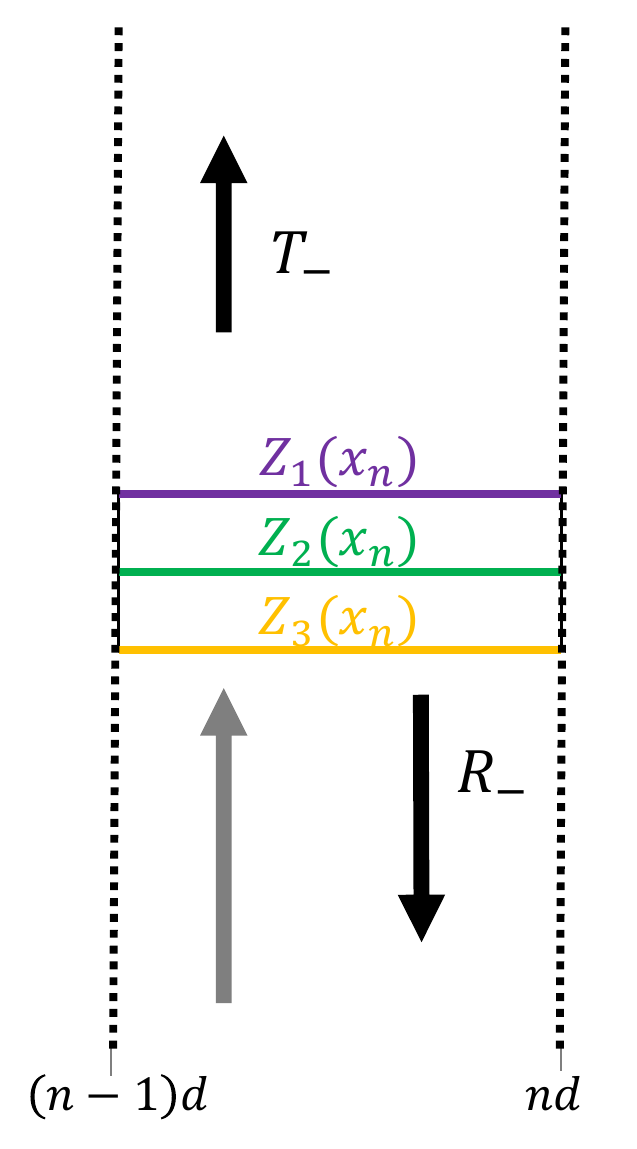}\label{fig:FIG9B}}
\end{minipage}
\begin{minipage}{0.35\columnwidth}
\subfigure[]{\includegraphics[width=1\linewidth]{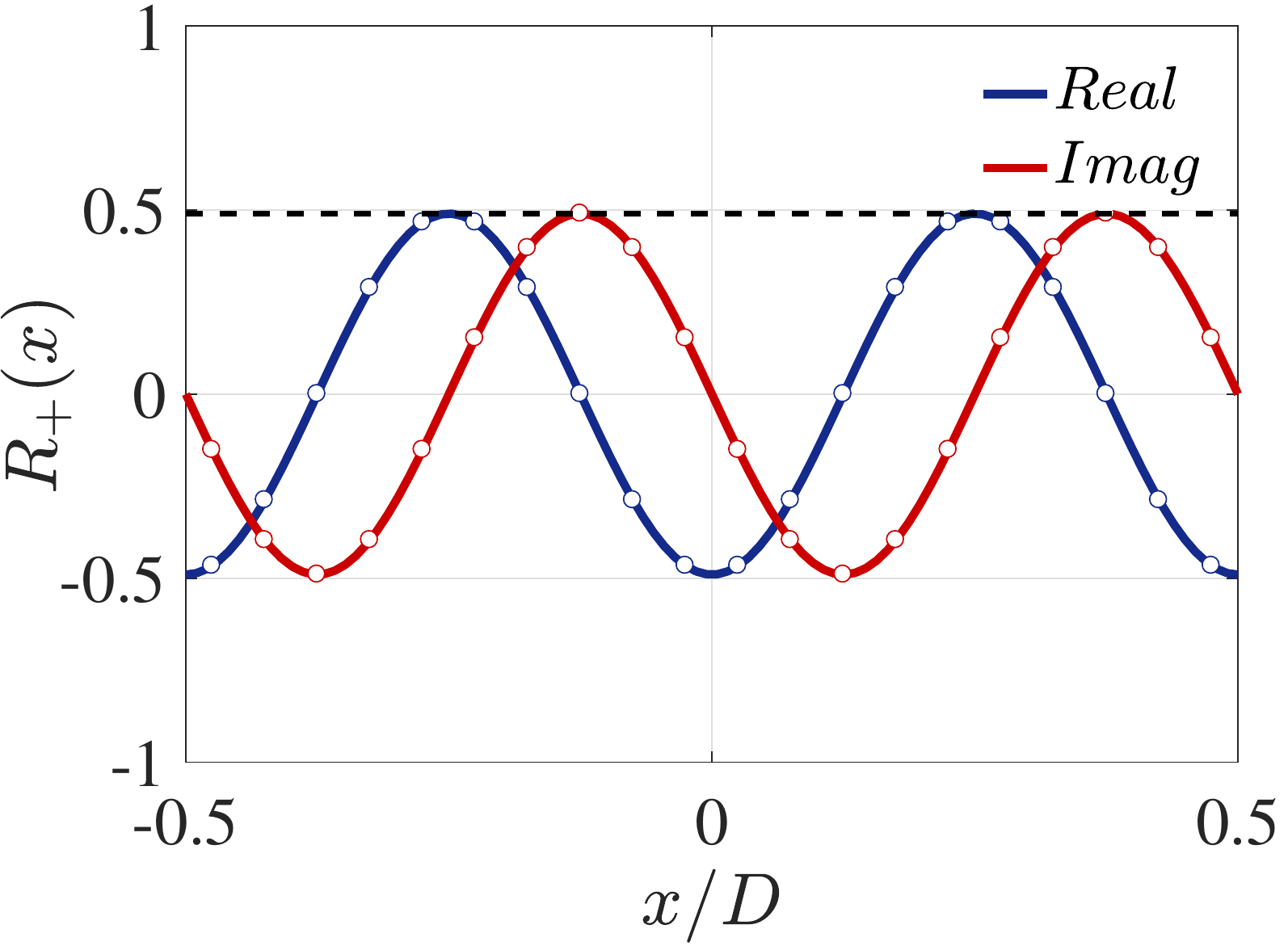}}
\subfigure[]{\includegraphics[width=1\linewidth]{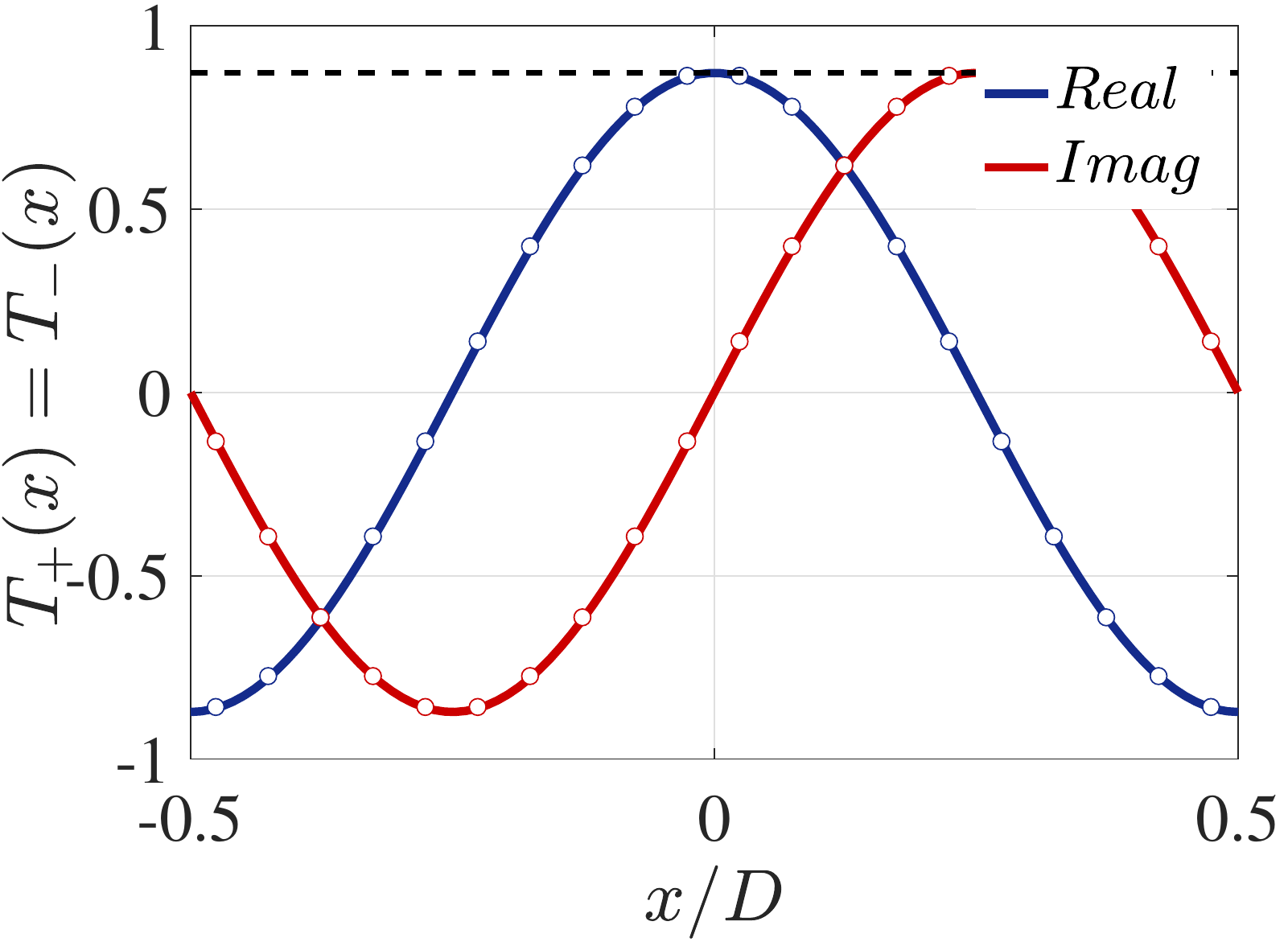}}
\subfigure[]{\includegraphics[width=1\linewidth]{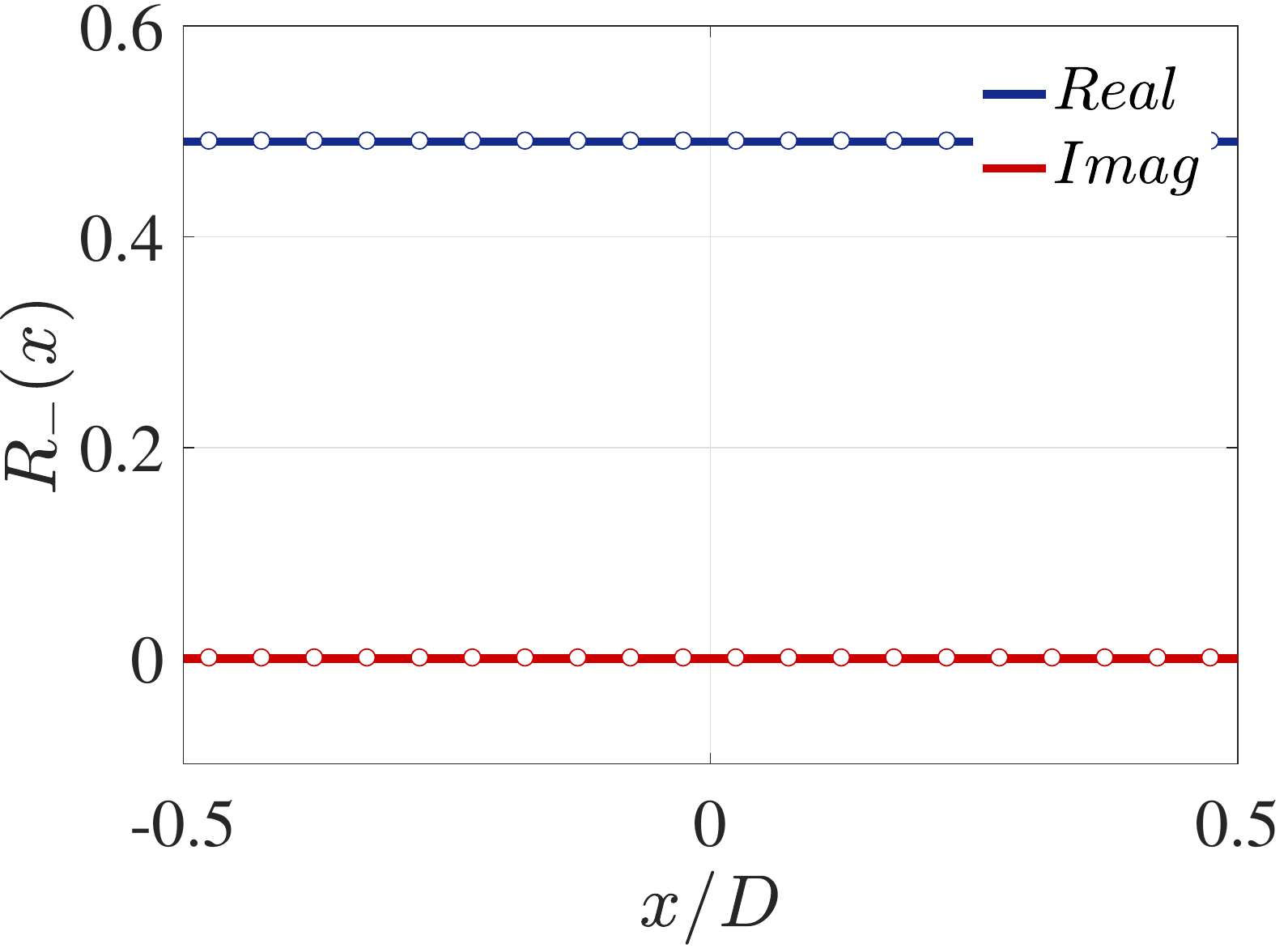}}
\end{minipage}
\caption{Schematic representation of the forward (a) and backward (b) illumination scenarios. Forward (+) and backward (-) local transmission and reflection coefficients for a perfect refractive metasurface. (c) Forward reflection, (d) forward  and backward transmission, and (e) backward reflection.  }\label{fig:S_parameters}
\end{figure}

It is also interesting to consider the local transmission and reflection coefficients in this scenario. Due to the asymmetry, the response of the particles when they are illuminated from media I (forward illumination) and II (backward illumination) will be different [see Fig. \ref{fig:FIG9A} and  Fig. \ref{fig:FIG9B}]. Transmission and reflection coefficients can be calculated using the impedance matrix as follows:
\begin{eqnarray}
R_{+}=\frac{(Z_{11}-Z_0)(Z_{22}+Z_0)-Z_{12}^2}{\Delta Z},\\
T_{+}=T_{-}=\frac{2Z_{12}Z_0}{\Delta Z},\\
R_{-}=\frac{(Z_{11}+Z_0)(Z_{22}-Z_0)-Z_{12}^2}{\Delta Z}, 
\end{eqnarray}
where $\Delta Z=(Z_{11}+Z_0)(Z_{22}+Z_0)-Z_{12}^2$ and the $\pm$ signs correspond to the forward and backward illuminations. Calculated results are plotted in Fig.~\ref{fig:S_parameters}. We can see how individually the meta-atoms are not matched and generate forward and backward reflections. 

\begin{figure}[h!]
	\centering
	\begin{minipage}{1\columnwidth}
		\subfigure[]{\includegraphics[height=5.5cm]{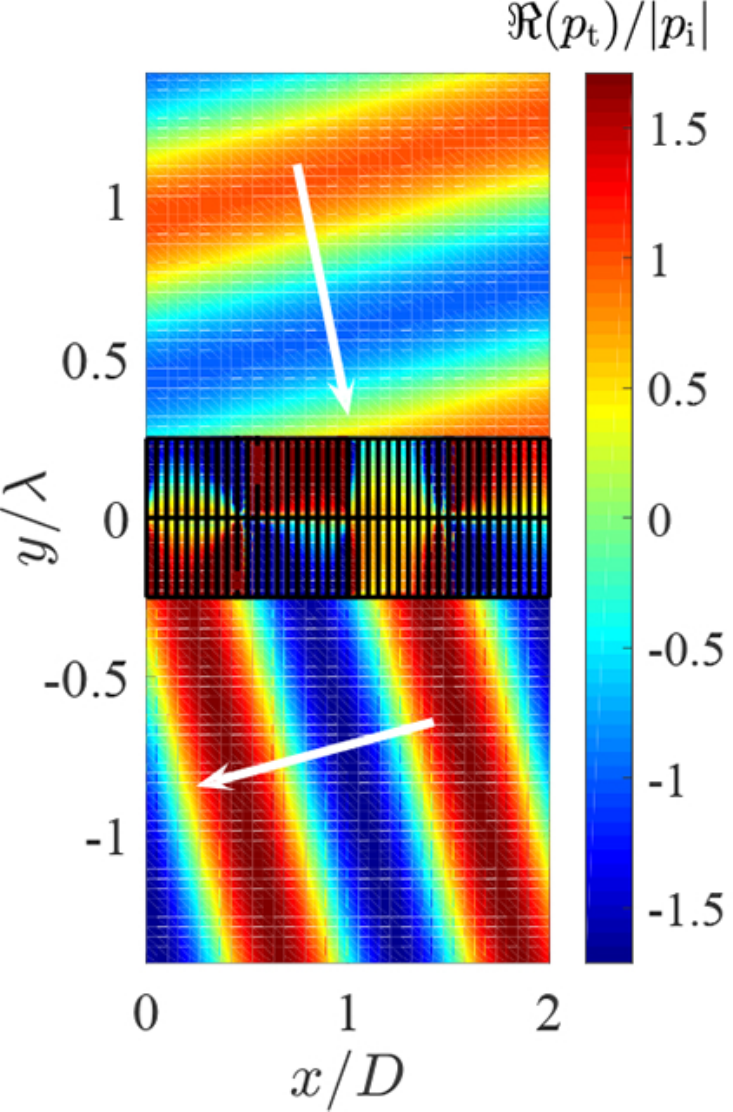}\label{fig:Tx_11_A}}
		\subfigure[]{\includegraphics[height=5.5cm]{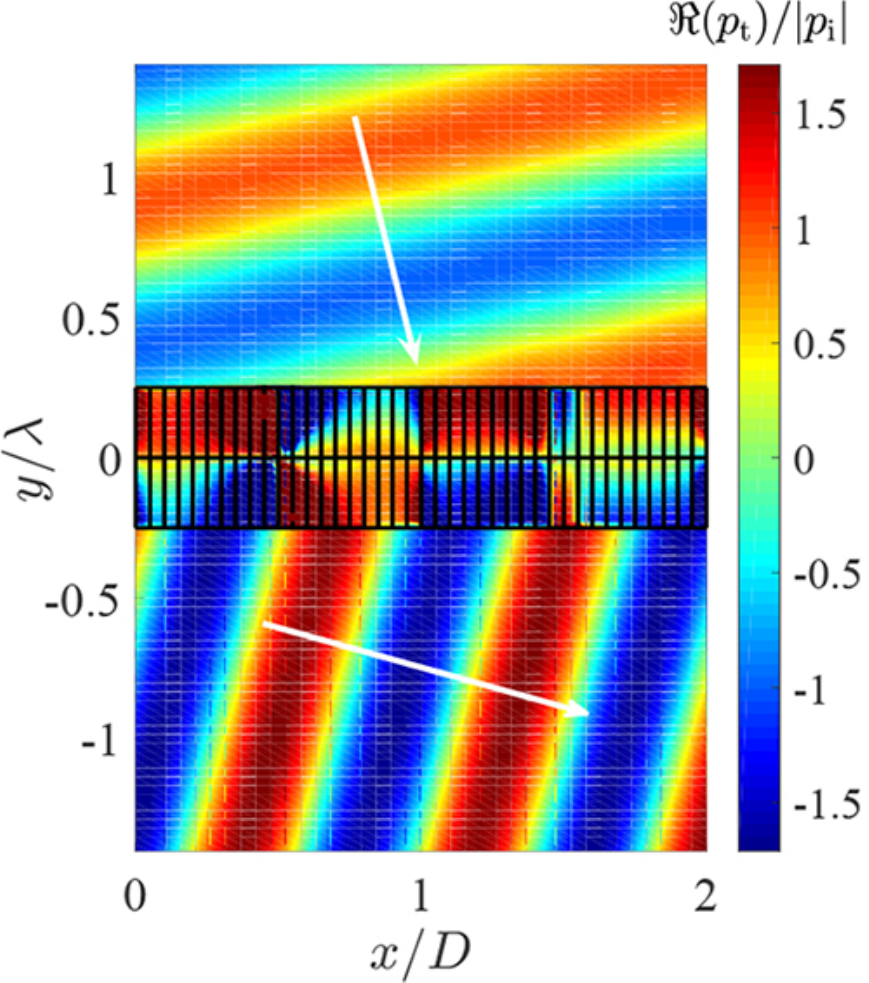}\label{fig:Tx_11_B}}
	\end{minipage}
	\begin{minipage}{1\columnwidth}
		\subfigure[]{\includegraphics[width=0.45\linewidth]{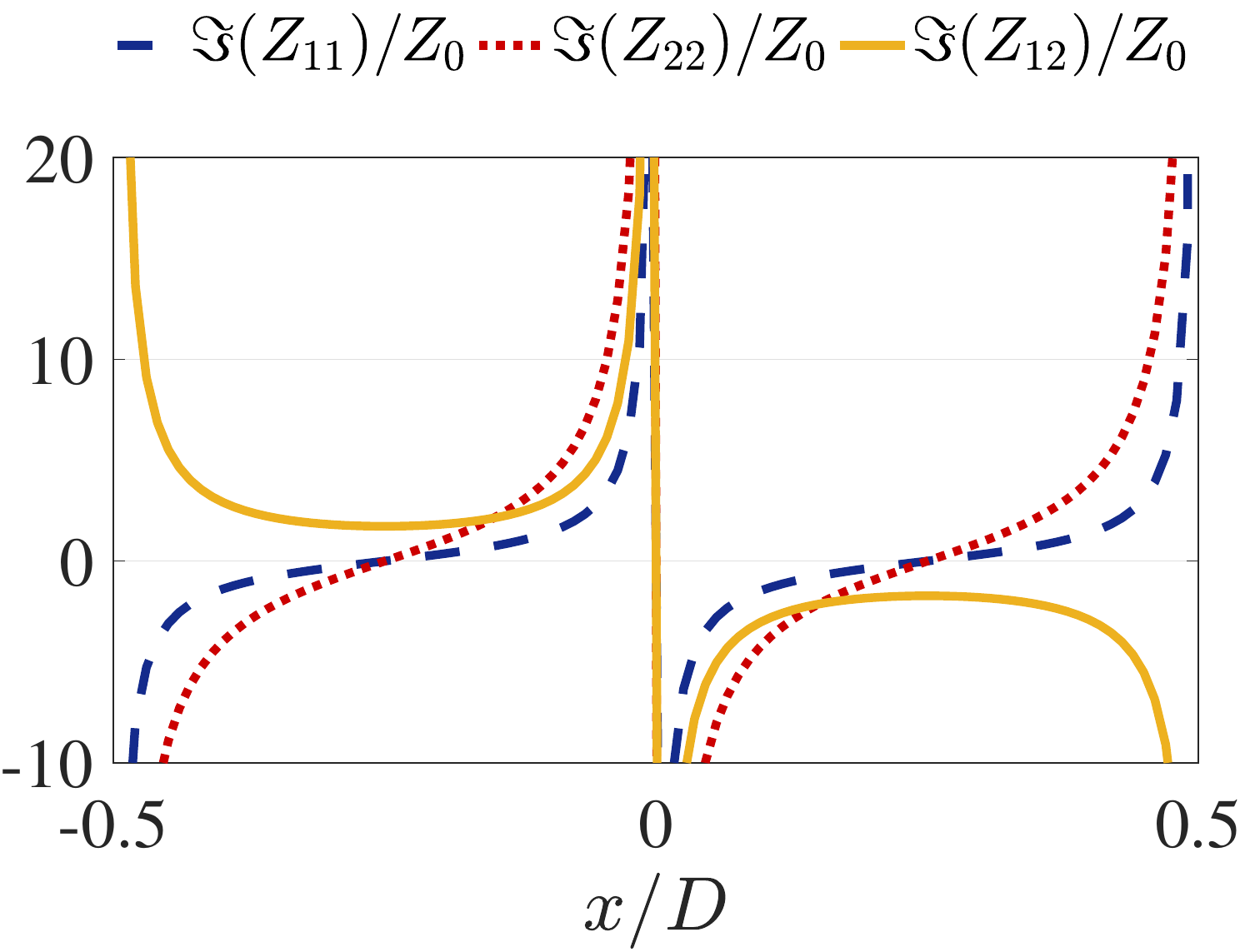}\label{fig:Tx_11_C}}
		\subfigure[]{\includegraphics[width=0.45\linewidth]{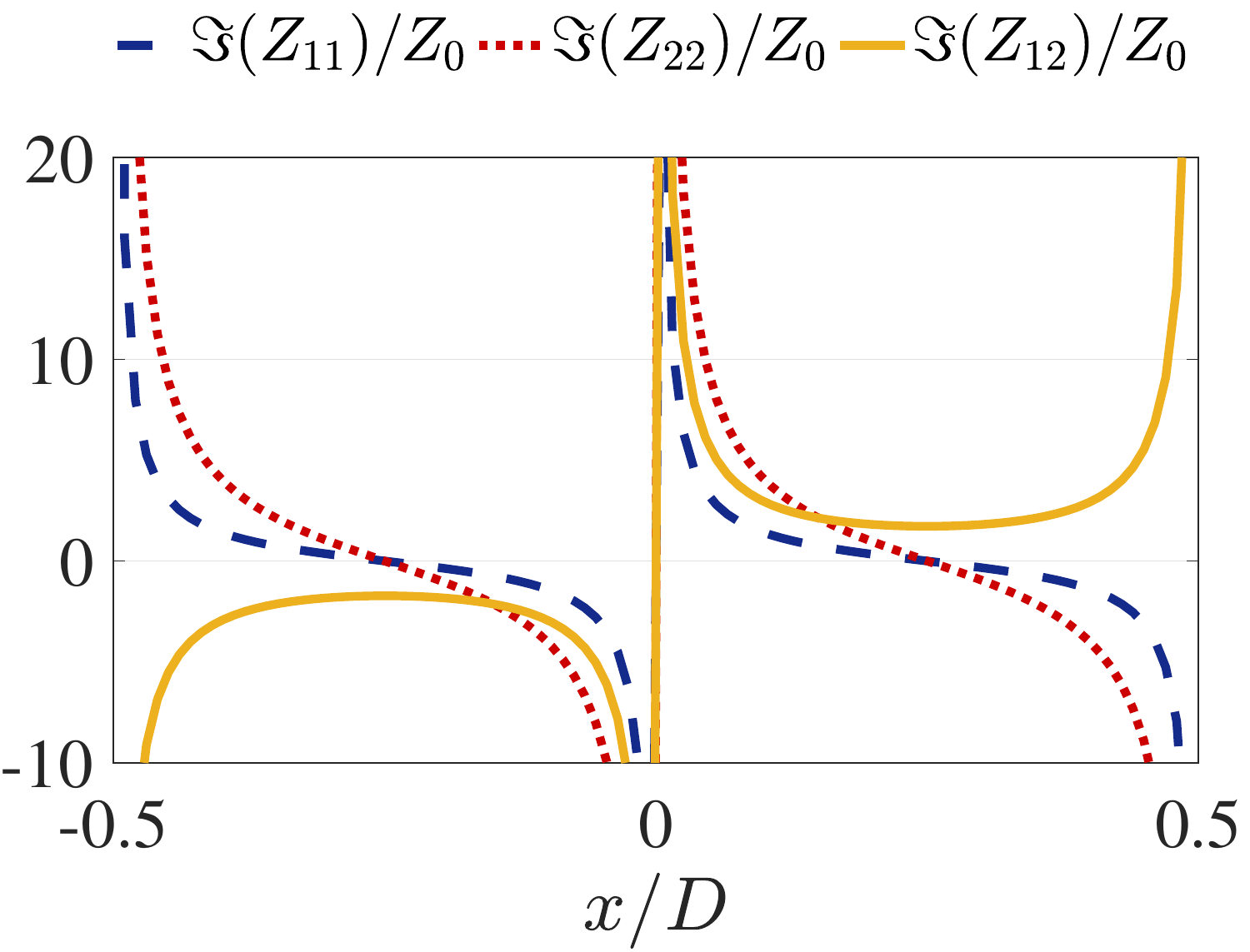}\label{fig:Tx_11_D}}
	\end{minipage}
	\caption{Anomalous refractive metasurface for  $\theta_{\rm i}=10^\circ$ and  $\theta_{\rm r}=- 70^\circ$: (a) Real part of the total pressure field for the 3-membranes implementation and (c) $Z$-matrix components. Anomalous refractive metasurface for  $\theta_{\rm i}=10^\circ$ and  $\theta_{\rm r}=70^\circ$: (b) Real part of the total pressure field for the 3-membranes implementation and (d) $Z$-matrix components.   }\label{fig:Tx_functionalities}
\end{figure}

The above  results demonstrate a possibility to change the refraction angle $\theta_{\rm r}$ when the metasurface is illuminated normally. The method can be easily applied for other incidence angles.  Figure~\ref{fig:Tx_functionalities} illustrates two different designs for a metasurface illuminated at $\theta_{\rm i}=10^\circ$. In the first scenario the incident plane wave is refracted into $\theta_{\rm r}=-70^\circ$. The relation between the amplitudes of the incident and reflected waves is defined by Eq.~(\ref{eq:Tx_amplitudes}) being $A=1.69$. The $Z$-matrix components of this structure are represented in Fig.~\ref{fig:Tx_11_C}. We can implement it by using the 3-membranes topology as it is shown in Fig. ~\ref{fig:Tx_11_A}. The second designed metasurface changes the direction of the transmitted wave from $\theta_{\rm i}=10^\circ$  to $\theta_{\rm r}=70^\circ$. The relation between the amplitudes of the incident and transmitted waves is the same as in the previous example,  and the $Z$-matrix components are represented in  Fig.~\ref{fig:Tx_11_D}. An important difference between both designs is the period of the metasurface which equals $D=\lambda/\vert \sin{\theta_{\rm r}} - \sin{\theta_{\rm i}} \vert$.   

\section{Conclusions}

This work introduces a new approach for the synthesis of acoustic metasurface for anomalous transmission and reflection.  We have explained the main ideas of the method by using a simple model based on the inhomogeneous surface impedance of the metasurface. This model has allowed direct comparison between conventional designs based on the generalized reflection and  Snell's laws and the introduced new approach, showing drastic improvements in power efficiency.  The fundamental advance introduced by our method is the full suppression of parasitic reflections in non-desired directions which reduce the total efficiency of the metasurface. 

Applying the introduced design  method to the reflection and transmission scenarios we have identified different physical phenomena which need to be realized for the ideal performance.  Metasurfaces for controlling reflection must exhibit non-local response, which allows energy channeling along the metasurface. In contrast, for full control of plane-wave transmission local but asymmetric response is required for each particle. 

Based on the general synthesis theory, we have identified appropriate topologies of acoustic unit cells which allow realization of perfect reflection and refraction. For perfect reflection, arrays of acoustical stubs can be used, and for perfect refraction, one can use three-membranes unit cells. To realize the required non-local properties of reflecting surfaces, the unit cells in each super-cell need to be optimized together, including near-field couplings between the cells. We have exemplified this procedure with the design of an acoustic anomalous reflector with 95\% of efficiency when $\theta_{\rm i}=0^\circ$ and $\theta_{\rm r}=70^\circ$.   In the transmission scenario, the required asymmetry of unit cells can be realized if all three membranes of each unit cell are different. Since the proposed synthesis method allows complete suppression of  parasitic reflection, the only factor which will limit the power efficiency is the power dissipation due to inevitable losses in the materials from which the metasurface is made. 

We hope that this work will motivate future experimental demonstrations of perfect anomalous reflective and refractive metasurfaces. To conclude, let us stress the importance of this advance for the improvement of other acoustic systems where  perfect performance in the sense of suppression of parasitic scattering is  desired. The proposed design methodology can be extended for arbitrary manipulations of multiple plane waves allowing more complex functionalities. In general, by designing amplitudes and phases of different waves and ensuring the local conservation of the power, it will be possible to overcome the efficiency drawbacks of the existing solutions for arbitrarily transformations of acoustic fields. 

\section*{Acknowledgment}
This work was supported in part by the Academy of Finland (projects 13287894 and 13309421).

\end{document}